\documentclass[showpacs,aps,prd,reprint,superscriptaddress,nofootinbib,longbibliography]{revtex4-1}
\usepackage[colorlinks=true, pdfstartview=FitV, linkcolor=magenta,citecolor=blue, urlcolor=blue,
bookmarks=true, bookmarksnumbered=true, breaklinks]{hyperref}
\usepackage[dvipdfmx]{graphicx}
\usepackage{amsmath,amssymb,bm,color,longtable,mathrsfs,slashed,comment,tikz}
\definecolor{VioletRed1}{rgb}{1,0.244,0.59}
\definecolor{DarkOrchid1}{rgb}{0.75,0.244,1}
\definecolor{SeaGreen4}{rgb}{0.18,0.545,0.34}
\definecolor{RoyalBlue3}{rgb}{0.228,0.372,0.804}
\hypersetup{linkcolor=VioletRed1, citecolor=DarkOrchid1, urlcolor=RoyalBlue3}

\definecolor{lime}{HTML}{A6CE39}
\DeclareRobustCommand{\orcidicon}{
	\hspace{-3mm}
	\begin{tikzpicture}
	\draw[lime, fill=lime] (0,0) 
	circle [radius=0.16] 
	node[white] {{\fontfamily{qag}\selectfont \tiny ID}};
	\draw[white, fill=white] (-0.0625,0.095) 
	circle [radius=0.007];
	\end{tikzpicture}
	\hspace{-3mm}
}

\foreach \x in {A, ..., Z}{\expandafter\xdef\csname orcid\x\endcsname{\noexpand\href{https://orcid.org/\csname orcidauthor\x\endcsname}
			{\noexpand\orcidicon}}
}

\begin{document}
\begin{flushright}
\end{flushright}

\title{Two relativistic Kondo effects: Classification with particle and antiparticle impurities}

\author{Yasufumi~Araki\orcidA{}}
\email[]{{araki.yasufumi@jaea.go.jp}}
\affiliation{Advanced Science Research Center, Japan Atomic Energy Agency (JAEA), Tokai 319-1195, Japan}

\author{Daiki~Suenaga\orcidB{}}
\email[]{suenaga@rcnp.osaka-u.ac.jp}
\affiliation{Research Center for Nuclear Physics (RCNP), Osaka University, Osaka 567-0047, Japan}

\author{Kei~Suzuki\orcidC{}}
\email[]{{k.suzuki.2010@th.phys.titech.ac.jp}}\thanks{corresponding author.}
\affiliation{Advanced Science Research Center, Japan Atomic Energy Agency (JAEA), Tokai 319-1195, Japan}

\author{Shigehiro~Yasui\orcidD{}}
\email[]{yasuis@keio.jp}
\affiliation{Research and Education Center for Natural Sciences, Keio University, Hiyoshi 4-1-1, Yokohama, Kanagawa 223-8521, Japan}
\affiliation{RIKEN iTHEMS, RIKEN, Wako 351-0198, Japan}

\date{\today}

\begin{abstract}
We investigate two different types of relativistic Kondo effects, distinguished by heavy-impurity degrees of freedom, by focusing on the energy-momentum dispersion relations of the ground state with condensates composed of a light Dirac fermion and a nonrelativistic impurity fermion.
Heavy fermion degrees of freedom are introduced in terms of two types of heavy-fermion effective theories, in other words, two heavy-fermion limits for the heavy Dirac fermion, which are known as the heavy-quark effective theories (HQETs) in high-energy physics.
While the first one includes only the heavy-particle component, the second one contains both the heavy-particle and heavy-antiparticle components, which are opposite in their parity.
From these theories, we obtain two types of Kondo effects, in which the dispersions near the Fermi surface are very similar, but they differ in the structure at low momentum.
We also classify the possible forms of condensates in the two limits.
The two Kondo effects will be examined by experiments with Dirac/Weyl semimetals or quark matter, lattice simulations, and cold-atom simulations.
\end{abstract}

\maketitle

\section{Introduction} \label{Sec:1}
The Kondo effect \cite{Kondo:1964,Hewson,Yosida,Yamada,coleman_2015} attracts much broad interest in the fields of both condensed-matter and high-energy physics.
While the original Kondo effect is physics for nonrelativistic electrons, studies of the Kondo systems with relativistic (Dirac/Weyl/Majorana) fermions will be useful for comprehensive understandings of Kondo effects realized in Dirac/Weyl metals and semimetals \cite{Principi:2015,Yanagisawa:2015conf,Yanagisawa:2015,Mitchell:2015,Sun:2015,Feng:2016,Kanazawa:2016ihl,Lai:2018,Ok:2017,PhysRevB.97.045148,PhysRevB.98.075110,PhysRevB.99.115109,KIM2019236,Grefe:2019} described by effective Hamiltonians, in nuclear matter~\cite{Yasui:2013xr,Yasui:2016ngy,Yasui:2016hlz,Yasui:2019ogk} described by hadronic effective models, and in quark matter~\cite{Yasui:2013xr,Hattori:2015hka,Ozaki:2015sya,Yasui:2016svc,Yasui:2016yet,Kanazawa:2016ihl,Kimura:2016zyv,Yasui:2017izi,Suzuki:2017gde,Yasui:2017bey,Kimura:2018vxj,Macias:2019vbl,Hattori:2019zig,Suenaga:2019car,Suenaga:2019jqu,Kanazawa:2020xje} described by the quantum chromodynamics (QCD) or effective models of QCD.

The Kondo effect is induced by the interplay between light fermions with a Fermi surface and heavy particles as impurities.
Theoretically, in order to precisely formulate the Kondo effect, one should be careful about which and how many degrees of freedom of heavy particles are participating in this interplay effect as impurities.
In this paper, we demonstrate that the difference in the Kondo effect arises from the choice of heavy-particle degrees of freedom, by taking two kinds of ``heavy-fermion limits" (or nonrelativistic limits) based on the heavy-quark effective theory (HQET)~\cite{Eichten:1989zv,Georgi:1990um} well known in high-energy particle physics (see Refs.~\cite{Neubert:1993mb,Manohar:2000dt} for reviews).
The first one is the {\it conventional HQET}~\cite{Mannel:1991mc}, where the heavy field contains only a single component, either the particle or antiparticle component.
The second one~\cite{Korner:1991kf,Balk:1993ev,Das:1993rf,Das:1993jx,Balk:1993cd,Blok:1996iz,Holstein:1997,Gardestig:2007mk} is based on the Foldy-Wouthuysen (FW) transformation~\cite{Foldy:1949wa,Tani:1951,Feinberg:1977rc} of the Dirac field, which is called the {\it FW-HQET} in Ref.~\cite{Balk:1993ev}.
In this theory, the heavy field is composed of the particle and antiparticle components.
Both of the HQETs are derived by a transformation of the original massive Dirac field and by the expansion with respect to the inverse heavy mass.

\begin{figure*}[tb!]
    \begin{minipage}[t]{0.66\columnwidth}
        \begin{center}
            \includegraphics[clip, width=1.0\columnwidth]{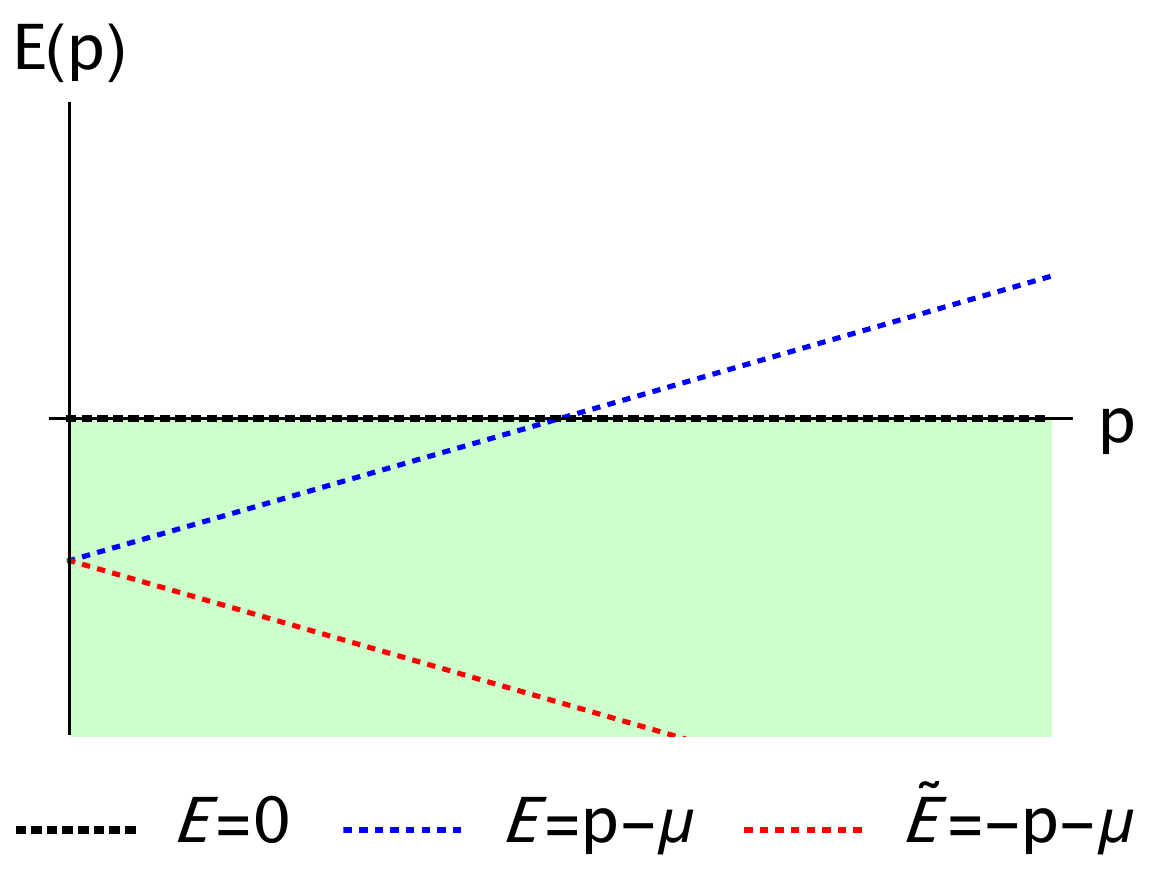}
        \end{center}
    \end{minipage}%
    \begin{minipage}[t]{0.66\columnwidth}
        \begin{center}
            \includegraphics[clip, width=1.0\columnwidth]{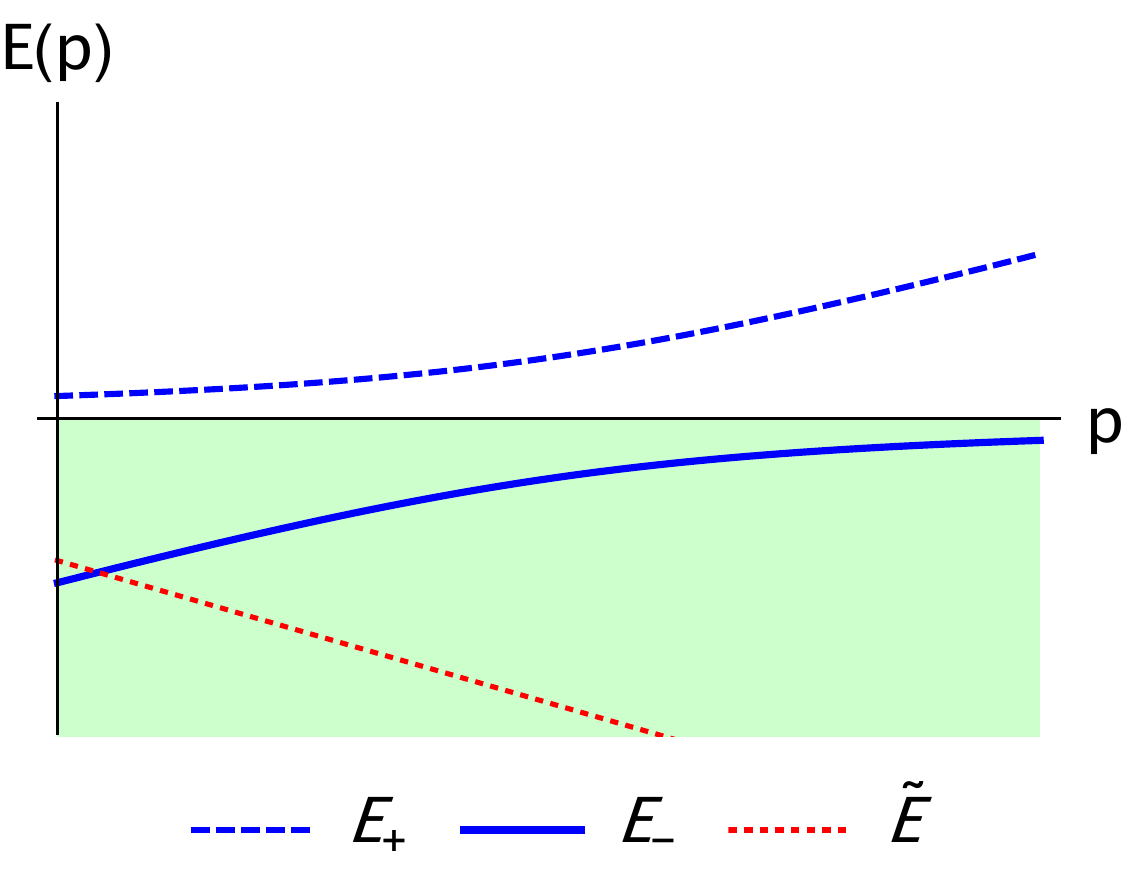}
        \end{center}
    \end{minipage}%
    \begin{minipage}[t]{0.66\columnwidth}
        \begin{center}
            \includegraphics[clip, width=1.0\columnwidth]{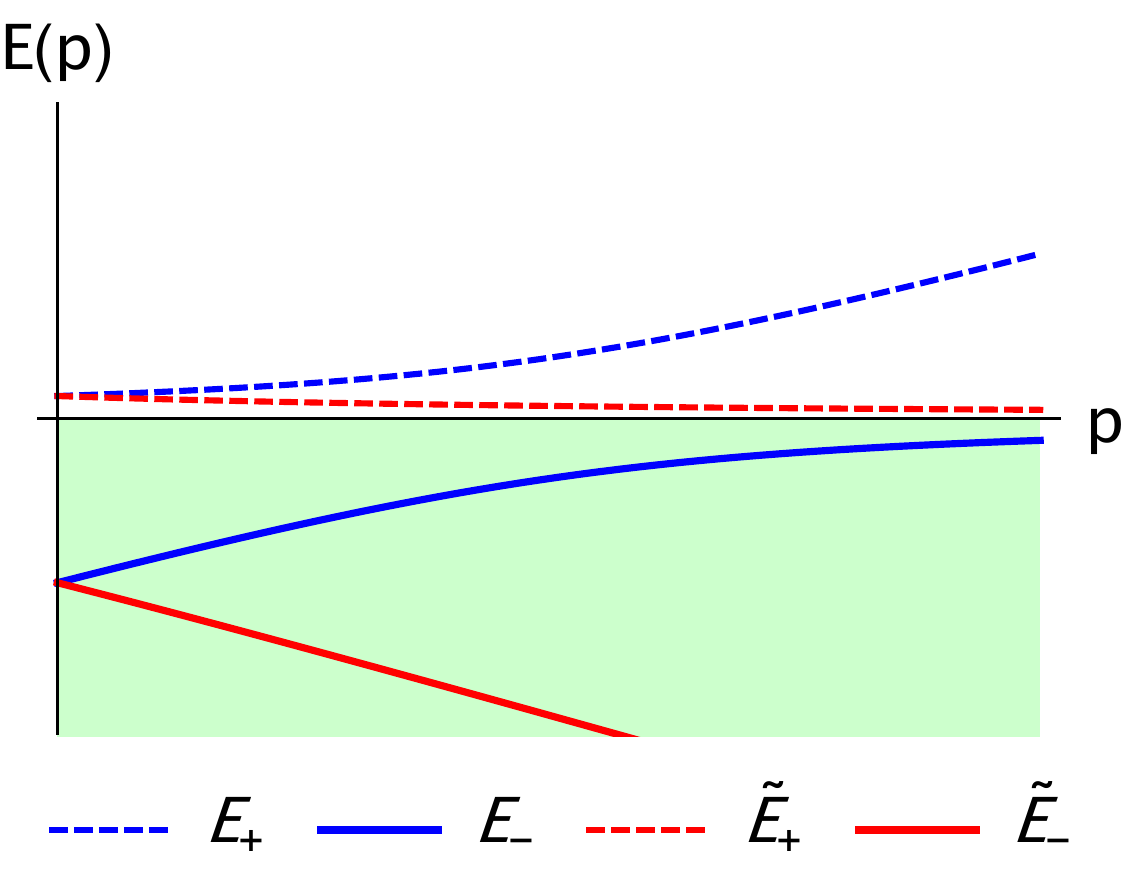}
        \end{center}
    \end{minipage}
    \caption{Examples of dispersion relations of quasiparticles with $S+V$ condensate at $\lambda=0$.
The colored region means the Fermi sea.
Left: Dispersion relations of light Dirac fermions ($E=\pm p- \mu$) and heavy fermions ($E=0$) without condensates.
Middle: Eqs.~(\ref{eq:Epm_SV_HQET}) and (\ref{eq:Etilde_SV_HQET}) in the conventional HQET.
Right: Eqs.~(\ref{eq:Epm_SV_FWHQET}) and (\ref{eq:Etilde_SV_FWHQET}) in the FW-HQET.}
\label{fig:disp}
\end{figure*}

We suggest that the difference between these two heavy-fermion limits leads to {\it similar but slightly different} properties of the relativistic Kondo effect, which are characterized as dispersion relations, as will be shown later in Fig.~\ref{fig:disp}.
Here, we emphasize the insights from these Kondo effects as follows:
\begin{itemize}
\item[(i)] {\it Similarity between two Kondo effects}---Their structures of dispersion near the Fermi surface are almost similar, where a hybridization between the light-particle branch and the heavy-particle branch is realized due to a nonzero condensate.
This similarity is based on the fact that the dominant degree of freedom in the Kondo effects is the heavy particle, and the contribution from the heavy antiparticle is relatively small.
\item[(ii)] {\it Difference between two Kondo effects}---On the other hand, a difference is found in the light-antiparticle branches in the low-momentum region away from the Fermi surface.
Light particle and antiparticle branches in the FW-HQET are simultaneously modified by a condensate, while, in the conventional HQET, the light-antiparticle branches can be decoupled from condensates. 
As a result, for the former we find that a pointlike (Dirac-cone-like) crossing structure between the light-particle and light-antiparticle branches remains at zero momentum.
For the latter, such a pointlike crossing can be broken.
\end{itemize}

Our analysis on the two Kondo effects will be useful for understanding various condensate (or hybridization) structures assumed in mean-field theories.
For example, in Refs.~\cite{Yasui:2016svc,Yasui:2017izi}, to nonperturbatively describe the relativistic Kondo effect by using a mean-field approximation, the authors proposed an ansatz of a condensate composed of a heavy particle and a light antiparticle with the scalar and vector-``hedgehog" Lorentz structure defined as $1+\hat{p} \cdot \vec{\gamma}$, where $\hat{p} \equiv \vec{p}/|\vec{p}|$ is the unit vector of the there-dimensional momentum, and $\vec{\gamma}\equiv (\gamma^1, \gamma^2, \gamma^3)$ is the spatial component of the Dirac gamma matrix.
They named it {\it Kondo condensate} (for applications of this condensate, see Refs.~\cite{Suzuki:2017gde,Yasui:2017bey,Macias:2019vbl,Suenaga:2019car,Suenaga:2019jqu}).
To the contrary, in Refs.~\cite{Feng:2016,Lai:2018}, the authors found a similar solution for Dirac/Weyl fermions {\it by assuming only the  condensate with the scalar-type Lorentz structure}.
In this paper, based on the two types of HQETs, we classify various {\it Ans\"atze} including other types of condensates.

In addition, a similar hedgehog-type hybridization was also suggested in the context of the nonrelativistic topological Kondo insulators (TKIs) such as SmB${}_{6}$ and YbB${}_{12}$~\cite{DzeroSunGalitskiColeman2010,DzeroSunColemanGalitski2012,TranTakimotoKim2012,Takimoto:2011,WernerAssaad2013,Alexandrov:2013,Werner:2014,Legner:2014,Roy:2014,Alexandrov:2015,Legner:2015,DzeroXiaGalitskiColeman2016,Baruselli:2016,Takasan:2017,Chang:2017,Peters:2018,Li_1806.05578,Li_1809.09867,Lu:2019,Peters:2019,Zhuang:2019,Tada:2020} (see Refs.~\cite{Alexandrov:2014,Lobos:2014,Mezio:2015,Hagymasi:2016,Zhong:2016,Lisandrini:2017,Zhong:2017,Pillay:2018,Hagymasi:2018,Zhong:2019} for one-dimensional TKIs).
In this paper, we compare hybridizations in TKIs and hedgehog condensates in relativistic Kondo effects.

This paper is organized as follows.
In Sec.~\ref{Sec:2}, we investigate the dispersion relations of quasiparticles in the relativistic Kondo effects based on the two types of HQETs.
Here, we classify the possible forms of condensates generating a simple form of dispersion relations.
In Sec.~\ref{Sec:3}, we summarize the possible forms of Kondo condensates and discuss the comparison between relativistic Kondo condensates and a hybridization in nonrelativistic TKIs.
Section~\ref{Sec:4} is devoted to our discussion and outlook.

Note that throughout this paper, we employ the Dirac representation for the $4 \times 4$ Dirac matrices:
\begin{eqnarray}
\gamma_0 = \left(
\begin{array}{cc}
1 & 0 \\
0 & -1 \\
\end{array}
\right), \ \ \vec{\gamma} = \left(
\begin{array}{cc}
0 & \vec{\sigma} \\
-\vec{\sigma} & 0 \\
\end{array}
\right), \ \ \gamma_5 = \left(
\begin{array}{cc}
0 & 1 \\
1 & 0 \\
\end{array}
\right),
\end{eqnarray}
where $\vec{\sigma}$ is the $2 \times2 $ Pauli matrix.

\section{Comparison of Kondo effects from two HQETs} \label{Sec:2}
In this section, we investigate the dispersion relations of quasiparticles for the ground states under relativistic Kondo effects.
The Kondo effect is induced by a non-Abelian interaction between a light fermion and a heavy fermion.
In order to formulate relativistic Kondo effects, we start from an effective Lagrangian with a massless Dirac field $\psi_1$ and massive Dirac field $\psi_2$,
\begin{align}
{\cal L}_\mathrm{eff}
=& \bar{\psi}_1 \bigl(i\slashed{\partial} +\mu\gamma_0\bigr)\psi_1 + \bar{\psi}_2 \bigl(i\slashed{\partial} - m - \lambda \gamma_0\bigr)\psi_2 
\nonumber \\
& + G\bigl( \bar{\psi}_1 \gamma_\mu T^a  \psi_1 \bigr) \bigl( \bar{\psi}_2 \gamma^\mu T^a  \psi_2 \bigr), \label{eq:2flavor_1}
\end{align}
in which an interaction between $\psi_1$ and $\psi_2$ is described by a four-point vector-current-type one, which is the so-called ``Nambu--Jona-Lashinio--type" interaction~\cite{Nambu:1961tp,Nambu:1961fr} when $\psi_1= \psi_2$.
$\mu$ and $\lambda$ are chemical potentials for $\psi_1$ and $\psi_2$, respectively, and $m$ is the mass of $\psi_2$ (the minus sign of $-\lambda$ is not important\footnote{Sometimes, $-\lambda$ is regarded as not the chemical potential but the Lagrange multiplier for heavy impurities~\cite{Yasui:2016svc,Yasui:2017izi}.}).
$T^a$ is the generator of the non-Abelian group SU$(N)$,\footnote{In the QCD Kondo effect~\cite{Yasui:2013xr,Hattori:2015hka}, the non-Abelian group is SU$(3)$.} and $G$ is a constant.\footnote{In general, the spatial and temporal components of four-point coupling constant can be different.
See Ref.~\cite{Hattori:2015hka} for discussion from a perturbative approach for the QCD Kondo effect.
In this paper, we assume that the temporal and spatial couplings are equal.}
With the help of the Fierz transformation, the effective Lagrangian~(\ref{eq:2flavor_1}) can be rewritten in a more convenient form as
\begin{align}
{\cal L}_\mathrm{eff}
=& \bar{\psi}_1 \bigl(i\slashed{\partial} +\mu\gamma_0\bigr)\psi_1 + \bar{\psi}_2 \bigl(i\slashed{\partial} - m - \lambda \gamma_0\bigr)\psi_2 
\nonumber \\
& +2G\Bigl[|\bar{\psi}_1\psi_2|^2 +|\bar{\psi}_1 i\gamma_5 \psi_2|^2  \nonumber\\
& - \frac{1}{2} |\bar{\psi}_1 \gamma_\mu \psi_2|^2 - \frac{1}{2} |\bar{\psi}_1 \gamma_\mu \gamma_5 \psi_2|^2\Bigr] + \cdots, \label{eq:2flavor}
\end{align}
in the $1/N$ expansion,\footnote{For the conventional Kondo effect in solid-state physics, mean-field approaches with the $1/N$ expansion have been successfully applied.
For early works, see Refs.~\cite{Read:1983,Coleman:1983}.
Note that, for relativistic Kondo effects, additional degrees of freedoms of light fermions, such as chirality and flavors, sometimes induce the overscreened Kondo effect and non-Fermi-liquid behavior \cite{Kanazawa:2016ihl}, which is analogous to the multichannel Kondo effect in nonrelativistic systems \cite{Nozieres:1980}.
In such situations, the standard mean-field approach may not be applicable.
} in which, for example, $|\bar{\psi}_1\psi_2|^2$ means $(\bar{\psi}_1\psi_2)(\bar{\psi}_2\psi_1)$.

In the following analyses, we keep the first flavor $\psi_1$ and consider the two kinds of the heavy-fermion limits for the second flavor $\psi_2$: the conventional HQET and the FW-HQET.
In both the heavy-fermion limits, the four-momentum $P^\mu$ of heavy fields is represented by $P^\mu \sim m v^\mu + p^\mu$, where $m$, $v^\mu$, and $p^\mu$ are the mass, four-velocity, and virtual (residual) momentum of the heavy fields, respectively, if the mass is sufficiently heavier than the residual momentum ($ m \gg p^\mu$).

\subsection{Conventional HQET} \label{Sec:2-1}
In the conventional HQET, the second flavor is transformed to
\begin{align}
\psi_2 \to \psi_2 = e^{- imv \cdot x} \left( \Psi_v^+ + \Psi_v^- \right),
\end{align}
where
\begin{align}
&\Psi_v^+ \equiv e^{imv \cdot x} \frac{1}{2} (1+ v^\mu \gamma_\mu) \psi_2, \nonumber\\
&\Psi_v^- \equiv e^{imv \cdot x} \frac{1}{2} (1- v^\mu \gamma_\mu) \psi_2.
\end{align}
The redefined $\Psi_v^+$ is the velocity-dependent heavy field, which is the so-called ``large" component.
$\Psi_v^-$ is the ``small" component and neglected at the leading order of $1/m$ expansion.
At the rest frame of the heavy field, $v^\mu=(1,0,0,0)$, we get $\psi_2 \to e^{-imt} \frac{1+ \gamma_0}{2} \psi_2$.
Thus, the nonrelativistic momentum $mv^0$ (and also the heavy mass $m$) of the heavy field is subtracted as a plane wave $e^{- imt}$.
Furthermore, the antiparticle components of the Dirac four-component spinor for the heavy flavor are completely decoupled.\footnote{On the other hand, if we regard the heavy antiparticle field $\Psi_v^-$ as the large component, then the particle component is decoupled.}
Then, the effective Lagrangian (\ref{eq:2flavor}) is transformed to
\begin{align}
 {\cal L}_\mathrm{eff} 
=& \bar{\psi}_1 \bigl(i\slashed{\partial} +\mu\gamma_0\bigr)\psi_1 + \bar{\Psi}_v^+ \bigl(i \partial_0 - \lambda \bigr)\Psi_v^+ 
\nonumber \\
& +G\Bigl[|\bar{\psi}_1\Psi_v^+|^2 +|\bar{\psi}_1 i\gamma_5 \Psi_v^+|^2 \nonumber \\
& + |\bar{\psi}_1 \vec{\gamma}\Psi_v^+|^2  +|\bar{\psi}_1 \vec{\gamma} \gamma_5 \Psi_v^+|^2\Bigr]. \label{eq:HQET_eff}
\end{align}
For the four-point interaction terms, only the terms including $\Psi_v^+$ survive since we have neglected $\Psi_v^-$ component.

Next, we apply the following mean-field approximations for the four-point interaction terms in Eq.~(\ref{eq:HQET_eff}): 
\begin{align}
G \langle \bar{\psi}_1 \Psi_v^+ \rangle  &\equiv \Delta_S, \\
G \langle \bar{\psi}_1 i\gamma_5 \Psi_v^+ \rangle  &\equiv \Delta_P, \\
G \langle \bar{\psi}_1 \vec{\gamma} \Psi_v^+  \rangle  &\equiv \Delta_V \hat{p}, \\
G \langle \bar{\psi}_1 \gamma _5 \vec{\gamma} \Psi_v^+ \rangle &\equiv \Delta_A \hat{p}.
\end{align}
These {\it Ans\"atze} are defined in momentum space, and $\hat{p} \equiv \vec{p}/p$ ($p \equiv |\vec{p}|$) is the unit vector of the three-dimensional momentum $\vec{p}$.
In this paper, we focus on the four types of condensates.
The gap $\Delta_i$ of the condensate is a complex number, which means the strength of the mixing between the light and heavy fermions.
The indices of gaps, $i=S,P,V,A$, denote the scalar ($S$), pseudoscalar ($P$), vector ($V$), and axial vector ($A$), respectively.
We call the third (vector) and fourth (axial vector) condensates with $\hat{p}$ the {\it hedgehog condensates}, where $\hat{p}$ was introduced as one of the simplest three-dimensional vector satisfying the spherical symmetry in three-momentum space.
It should be noted that this {\it Ansatz} is not necessarily the unique solution of the ground state.
The shape of dispersion at low momentum may vary depending on the radial structure of the condensate in momentum space.
After assuming the hedgehog {\it Ans\"atze}, $\hat{p}$ is combined with another $\vec{\gamma}$ in the vector four-point interaction, and then we get the factor $\hat{p}\cdot \vec{\gamma}$.

As a result, the mean-field Lagrangian in the conventional HQET is\footnote{Eq.~(\ref{eq:HQET_MF_eff}) is the notation in momentum space.
If we write Eq.~(\ref{eq:HQET_MF_eff}) in real space, we should rewrite $\hat{p}\cdot \vec{\gamma}$ as $- i \hat{\nabla} \cdot \vec{\gamma}$, where $-i \hat{\nabla} \equiv -i \nabla/|-i\nabla|$ using the coordinate derivative $\nabla$.}
\begin{align}
{\cal L}_\mathrm{MF}
\equiv& \bar{\psi}_1 \bigl(\slashed{p} +\mu\gamma_0\bigr)\psi_1 + \bar{\Psi}_v^+ \bigl(p_0 - \lambda \bigr)\Psi_v^+ 
\nonumber \\
& +  \bar{\Psi}_v^+ (\Delta_S + i \Delta_P \gamma_5 + \Delta_V \hat{p}\cdot \vec{\gamma} + \Delta_A \hat{p}\cdot \vec{\gamma} \gamma_5) \psi_1 \nonumber \\
&+  \bar{\psi}_1 (\Delta^\ast_S + i \Delta^\ast_P \gamma_5 + \Delta^\ast_V \hat{p}\cdot \vec{\gamma} + \Delta^\ast_A \hat{p}\cdot \vec{\gamma} \gamma_5) \Psi_v^+  \nonumber \\
& -\frac{1}{G} \left( |\Delta_S|^2 + |\Delta_P|^2 + |\Delta_V|^2 + |\Delta_A|^2 \right),  \label{eq:HQET_MF_eff}
\end{align}
where we used $\gamma_0 \Psi_v^+ = \Psi_v^+$.
The inverse propagator of the ``quasiparticles" with all the $S$, $P$, $V$, and $A$ condensates is
\begin{align}
& G_\mathrm{MF}^{S,P,V,A}(p_{0},\vec{p}\,)^{-1}
\equiv \nonumber\\
& \left(
\begin{array}{ccc}
p_0 + \mu                        & -\vec{p}\cdot \vec{\sigma}& \Delta_S^\ast - \Delta_A^\ast \hat{p}\cdot \vec{\sigma}  \\
\vec{p}\cdot \vec{\sigma} & -(p_0 + \mu) & -\Delta_V^\ast \hat{p}\cdot \vec{\sigma} +i\Delta^\ast_P  \\
\Delta_S -\Delta_A \hat{p}\cdot \vec{\sigma}  & \Delta_V \hat{p}\cdot \vec{\sigma} +i\Delta_P & p_0 - \lambda     \\
\end{array}
\right).
\label{eq:Ginverse_SPVA_HQET}
\end{align}
Notice that this propagator consists of $6 \times 6$ components: the four-component Dirac spinor for the light flavor and the two-component spinor for the heavy flavor.
By assuming a combination of $S$, $P$, $V$, and $A$ condensates and by solving $\det[G(p_0)^{-1}]=0$, we can extract the various solutions for dispersion relations of quasiparticles.
In most cases, it is difficult to get a simple form of dispersion relations, but we can obtain a simple form only in a few examples such as $S+V$ and $V+A$.

As an example, we consider the case with $\Delta \equiv \Delta_S = \Delta_V$ as suggested in Ref.~\cite{Yasui:2016svc,Yasui:2017izi}.
For the inverse propagator with only the $S+V$ condensate, by solving $\det[G(p_0)^{-1}]=0$, we get the six energy-momentum dispersion relations
\begin{align}
E_\pm(p) &\equiv  \frac{1}{2} \left( p + \lambda -\mu \pm \sqrt{\left(p-\lambda-\mu \right)^2 + 8 |\Delta|^2 }\right) \nonumber\\
& \hspace{145pt} (\mathrm{Double}), \label{eq:Epm_SV_HQET} \\
\tilde{E}(p) &\equiv  - p - \mu \ \ \ (\mathrm{Double}). \label{eq:Etilde_SV_HQET}
\end{align}
These solutions are the same as those obtained in Ref.~\cite{Yasui:2016svc,Yasui:2017izi}.
$E_\pm(p)$ and $\tilde{E}(p)$ are doubly degenerate.

The antiparticle mode $\tilde{E}(p)$ is decoupled from the condensate.
This is a characteristic of the Kondo condensate in the conventional HQET, which is different from the FW-HQET as discussed below.
Note that the factor of $8$ in front of $|\Delta|^2$ might be characteristic in the sense that the factor of $2$ among $8$ stems from the structure of the ``combinatorial" condensate (or linear combination) of $S+V$.

In the middle panel of Fig.~\ref{fig:disp}, we plot an example of the dispersion relations, (\ref{eq:Epm_SV_HQET}) and (\ref{eq:Etilde_SV_HQET}).
Here, we can find that while the particle component of the light flavor is mixed with the heavy flavor as $E_\pm(p)$, the antiparticle component $\tilde{E}(p)$ is decoupled.
By a nonzero condensate, the original crossing-point (Dirac-point) structure between the light particle and light antiparticle near the origin ($p=0$) disappears, and as a result the crossing between $\tilde{E}(p)$ and $E_-(p)$ is located on the spherical surface in three-momentum space ($p \neq 0$, where remember $p \equiv |\vec{p}|$), which is no longer pointlike.
In the next subsection~\ref{Sec:2-2}, we will show that this situation changes in dispersion relations based on the FW-HQET. 

As another example, we assume $\Delta \equiv \Delta_V = \Delta_A$ and consider the case with the $V+A$ condensate.
From the inverse propagator with only the $V+A$ condensate, we get
\begin{align}
E_\pm(p) &\equiv  \frac{1}{2} \left( p + \lambda -\mu \pm \sqrt{\left(p-\lambda-\mu \right)^2 + 8 |\Delta|^2 }\right) \nonumber\\
& \hspace{145pt} (\mathrm{Single}), \\
\tilde{E}_\pm (p) &\equiv \frac{1}{2} \left( - p + \lambda -\mu \pm \sqrt{\left(p+\lambda+\mu \right)^2 + 8 |\Delta|^2 }\right) \nonumber\\
& \hspace{145pt} (\mathrm{Single}), \\
E(p)             &\equiv p-\mu \ \ \ (\mathrm{Single}), \\
\tilde{E}(p)     &\equiv -p-\mu \ \ \ (\mathrm{Single}).
\end{align}
These solutions are found in this paper.
Contrary to the $S+V$ condensate, these modes are not degenerate with each other, and both the particle $E(p)$ and antiparticle $\tilde{E}(p)$ modes are decoupled from the condensate.
Furthermore, at the origin $p=0$ in momentum space, we find that the Weyl-point structure between the decoupled modes, $E(p)$ and $\tilde{E}(p)$, and a crossing-point structure between the mixed modes, $E_-(p)$ and $\tilde{E}_-(p)$.

To summarize, in the conventional HQET, only for $S+V$ and $V+A$ condensates, we can get a simple form of the dispersion relations, as far as we checked.
For other condensates, we cannot obtain a simple form of dispersion relations after the diagonalization of the inverse propagators.
Therefore, in order to investigate the properties of the dispersion relation with these condensates, we have to numerically solve it.

\subsection{FW-HQET} \label{Sec:2-2}
In the previous subsection~\ref{Sec:2-1}, we studied the Kondo effects with the heavy-fermion field based on the conventional HQET, where only the particle component of the heavy-fermion field is included, and the antiparticle component is removed.
In this subsection, we introduce another type of heavy-fermion field based on the FW-HQET, where both the particle and antiparticle components are included.
Thus, an essential difference between the two types of HQETs is the existence or absence of the heavy-antiparticle component.

In the FW-HQET, the second flavor is transformed to~\cite{Korner:1991kf,Balk:1993ev}
\begin{align}
\psi_2 \to \psi_2 =& e^{- imv \cdot x} \left( e^{\frac{\mathcal{O}_1}{2m}} e^{\frac{\mathcal{O}_2}{2m^2}} e^{\frac{\mathcal{O}_3}{2m^3}} \cdots \right)  \Psi_v^+ \nonumber\\
&+  e^{imv \cdot x} \left( e^{\frac{\mathcal{O}_1}{2m}} e^{\frac{\mathcal{O}_2}{2m^2}} e^{\frac{\mathcal{O}_3}{2m^3}} \cdots \right) \Psi_v^-.
\end{align}
The $\Psi_v^+$ and $\Psi_v^-$ are defined as
\begin{align}
&\Psi_v^+ \equiv  \left( \cdots e^{-\frac{\mathcal{O}_3}{2m^3}} e^{-\frac{\mathcal{O}_2}{2m^2}} e^{-\frac{\mathcal{O}_1}{2m}} \right) e^{i mv \cdot x} \frac{1}{2}(1+ v^\mu \gamma_\mu) \psi_2, \\
&\Psi_v^- \equiv \left( \cdots e^{-\frac{\mathcal{O}_3}{2m^3}} e^{-\frac{\mathcal{O}_2}{2m^2}} e^{-\frac{\mathcal{O}_1}{2m}} \right) e^{-i m v \cdot x}  \frac{1}{2}(1- v^\mu \gamma_\mu) \psi_2,
\end{align}
where the differential operators $\mathcal{O}_i$ are determined order by order in the $1/m$ expansion of the massive Dirac field (see Refs.~\cite{Korner:1991kf,Balk:1993ev} for the explicit forms).
If we take the rest frame of the heavy field, $v^\mu=(1,0,0,0)$, we can use $ e^{\pm i mv \cdot x} \to e^{\pm i m t} $ and $\frac{1}{2}(1 \pm v^\mu \gamma_\mu) \to \frac{1\pm\gamma_0}{2}$.
$\frac{1 + \gamma_0}{2}$ and $\frac{1 - \gamma_0}{2}$ are the projection operators of the particle and antiparticle, respectively, and then $\Psi_v^+$ and $\Psi_v^-$ correspond to the heavy particle and heavy antiparticle, respectively.
Thus, by using the FW transformation in the rest frame, although the particle and antiparticle components are completely decoupled, both the fields can exist at the same time, which is definitely different from the conventional HQET.
Then, the effective Lagrangian (\ref{eq:2flavor}) is transformed to
\begin{align}
{\cal L}_\mathrm{eff} 
\equiv& \bar{\psi}_1 \bigl(i\slashed{\partial} +\mu\gamma_0\bigr)\psi_1 \nonumber \\
& + \bar{\Psi}_v^+ \bigl(i \partial_0 - \lambda_+ \bigr)\Psi_v^+ -\bar{\Psi}_v^- \bigl(i \partial_0 - \lambda_- \bigr)\Psi_v^- 
\nonumber \\
& +G\Bigl[|\bar{\psi}_1\Psi_v^+|^2 +|\bar{\psi}_1 i\gamma_5 \Psi_v^+|^2 \nonumber \\
& + |\bar{\psi}_1 \vec{\gamma}\Psi_v^+|^2  +|\bar{\psi}_1 \vec{\gamma} \gamma_5 \Psi_v^+|^2 \Bigr. \nonumber\\
\Bigl. &+|\bar{\psi}_1\Psi_v^-|^2 +|\bar{\psi}_1 i\gamma_5 \Psi_v^-|^2 \nonumber \\
&+ |\bar{\psi}_1 \vec{\gamma}\Psi_v^-|^2  +|\bar{\psi}_1 \vec{\gamma} \gamma_5 \Psi_v^-|^2 \Bigr]. \label{eq:FWHQET_eff}
\end{align}
For the four-point interaction terms, the terms including both $\Psi^+$ and $\Psi^-$, which are proportional to a rapidly oscillating phase factor ${\rm e}^{\pm2 imt}$, can be obtained. However, as long as $m$ is extremely larger than the typical energy scale such as the momentum or the strength of condensates, such terms will vanish.

Next, we apply the following mean-field approximations for the four-point interaction terms in Eq.~(\ref{eq:FWHQET_eff}):
\begin{align}
&G \langle \bar{\psi}_1 \Psi_v^+ \rangle \equiv \Delta_S^+,& &G \langle \bar{\psi}_1 \Psi_v^- \rangle \equiv \Delta_S^-,\\
&G \langle \bar{\psi}_1 i\gamma_5 \Psi_v^+ \rangle \equiv \Delta_P^+,& &G \langle \bar{\psi}_1 i\gamma_5 \Psi_v^- \rangle \equiv \Delta_P^-,\\
&G \langle \bar{\psi}_1 \vec{\gamma} \Psi_v^+  \rangle \equiv \Delta_V^+ \hat{p},& &G \langle \bar{\psi}_1 \vec{\gamma} \Psi_v^-  \rangle \equiv \Delta_V^- \hat{p},\\
&G \langle \bar{\psi}_1 \gamma _5 \vec{\gamma} \Psi_v^+ \rangle \equiv \Delta_A^+ \hat{p},& &G \langle \bar{\psi}_1 \gamma _5 \vec{\gamma} \Psi_v^- \rangle \equiv \Delta_A^- \hat{p},
\end{align}
Here, we assumed the condensates with different gaps, $\Delta^+_i$ and $\Delta^-_i$ ($i=S,P,V,A$), between the $\Psi_v^+$ and $\Psi_v^-$ fields.

As a result, the mean-field Lagrangian in the FW-HQET is
\begin{align}
{\cal L}_\mathrm{MF} 
\equiv& \bar{\psi}_1 \bigl(\slashed{p} + \mu\gamma_0\bigr)\psi_1 \nonumber \\
& + \bar{\Psi}_v^+ \bigl(p_0 - \lambda_+ \bigr)\Psi_v^+ -\bar{\Psi}_v^- \bigl(p_0 - \lambda_- \bigr)\Psi_v^- 
\nonumber \\
& +  \bar{\Psi}_v^+ (\Delta_S^+ + i \Delta_P^+ \gamma_5 + \Delta_V^+ \hat{p}\cdot \vec{\gamma} + \Delta_A^+ \hat{p}\cdot \vec{\gamma} \gamma_5) \psi_1 \nonumber \\
&+  \bar{\psi}_1 (\Delta^{+\ast}_S + i \Delta^{+\ast}_P \gamma_5 + \Delta^{+\ast}_V \hat{p}\cdot \vec{\gamma} + \Delta^{+\ast}_A \hat{p}\cdot \vec{\gamma} \gamma_5) \Psi_v^+  \nonumber \\
&+ \bar{\Psi}_v^- (\Delta^-_S + i \Delta^-_P \gamma_5 + \Delta^-_V \hat{p}\cdot \vec{\gamma} + \Delta^-_A \hat{p}\cdot \vec{\gamma} \gamma_5) \psi_1 \nonumber \\
&+  \bar{\psi}_1 (\Delta^{-\ast}_S + i \Delta^{-\ast}_P \gamma_5 + \Delta^{-\ast}_V \hat{p}\cdot \vec{\gamma} + \Delta^{-\ast}_A \hat{p}\cdot \vec{\gamma} \gamma_5) \Psi_v^-  \nonumber \\
& -\frac{1}{G} \left( |\Delta_S^+|^2 + |\Delta_P^+|^2 + |\Delta_V^+|^2 + |\Delta_A^+|^2 \right. \nonumber \\
& \left. + |\Delta_S^-|^2 + |\Delta_P^-|^2 + |\Delta_V^-|^2 + |\Delta_A^-|^2 \right), \label{eq:MFLag_SPVA_FW}
\end{align}
where we used $\gamma_0 \Psi_v^+ = \Psi_v^+$ and $\gamma_0 \Psi_v^- = -\Psi_v^-$.
The inverse propagator of quasiparticles is
\begin{widetext}
\begin{eqnarray}
 G_\mathrm{MF}^{S,P,V,A}(p_{0},\vec{p}\,)^{-1}
\equiv
 \left(
\begin{array}{cccc}
p_0 + \mu                        & -\vec{p}\cdot \vec{\sigma}& \Delta_{S}^{+\ast} - \Delta_A^{+\ast} \hat{p}\cdot \vec{\sigma} & \Delta_V^{-\ast} \hat{p}\cdot \vec{\sigma} +i\Delta^{-\ast}_P \\
\vec{p}\cdot \vec{\sigma} & -(p_0 + \mu) & -\Delta_V^{+\ast} \hat{p}\cdot \vec{\sigma} +i\Delta^{+\ast}_P & \Delta_S^{-\ast} + \Delta_A^{-\ast} \hat{p}\cdot \vec{\sigma} \\
\Delta_S^+ -\Delta_A^+ \hat{p}\cdot \vec{\sigma}  & \Delta_V^+ \hat{p}\cdot \vec{\sigma} +i\Delta_P^+ & p_0 - \lambda_+        & 0 \\
-\Delta_V^- \hat{p}\cdot \vec{\sigma}+i\Delta_P^- & \Delta_S^- +\Delta_A^- \hat{p}\cdot \vec{\sigma} & 0  & -(p_0 - \lambda_-) \\
\end{array}
\right). \label{eq:Ginverse_SPVA_FW}
\end{eqnarray}
\end{widetext}
This propagator consists of $8 \times 8$ components: the four-component Dirac spinor for the light flavor and the four-component spinor for the heavy flavor, which is definitely different from Eq.~(\ref{eq:Ginverse_SPVA_HQET}) with $6 \times 6$ components based on the conventional HQET.
In Eq.~(\ref{eq:Ginverse_SPVA_FW}), the off-diagonal components between the heavy-particle and heavy-antiparticle components are zero (in other words, $\vec{p}\cdot \vec{\sigma}$ is absent), which indicates that the heavy particle and heavy antiparticle are decoupled within the block matrix.
This decoupling is the consequence of the FW transformation.

We note that in the FW-HQET without the gaps, the heavy particle and antiparticle numbers are separately conserved~\cite{Balk:1993ev}.
The parameters $\lambda_\pm$ determine the relative positions of the dispersions of the heavy fermions from the Fermi surface of light fermions (determined by $\mu$), at the same time, and they serve as the chemical potentials for the numbers of heavy particles and antiparticles, measured in the residual momentum space.
For example, the condition of $\lambda_+ = \lambda_-$ means that the dispersions of the heavy particles and antiparticles coincide with each other, and the corresponding physical system contains equal numbers of heavy particles and antiparticles as long as we switch off the interaction between the light and heavy fermions.
Practically, for analyses of relativistic Kondo effects, it is enough to consider the condition of $\lambda_+ \approx \lambda_- \approx 0$ (for the role of $\lambda$ in the conventional HQET, see Refs.~\cite{Yasui:2016svc,Yasui:2017bey,Yasui:2017izi,Suzuki:2017gde,Macias:2019vbl,Suenaga:2019car,Suenaga:2019jqu}).

Next we extract the dispersion relations.
In order to obtain a simple solution, we assume $\Delta \equiv \Delta_S^+ = \Delta_V^+ = \Delta_S^- = \Delta_V^-$ and $\lambda \equiv \lambda_+ = \lambda_-$.
For the inverse propagator with only the $S+V$ condensate, by solving $\det[G(p_0)^{-1}]=0$, we get the dispersion relations
\begin{align}
E_\pm(p) &\equiv  \frac{1}{2} \left( p + \lambda -\mu \pm \sqrt{\left(p-\lambda-\mu \right)^2 + 8 |\Delta|^2 }\right) \nonumber\\
&\hspace{145pt} (\mathrm{Double}), \label{eq:Epm_SV_FWHQET}\\
\tilde{E}_\pm(p) &\equiv \frac{1}{2} \left( -p + \lambda -\mu \pm \sqrt{\left(p+\lambda+\mu \right)^2 + 8 |\Delta|^2 }\right) \nonumber\\
&\hspace{145pt} (\mathrm{Double}). \label{eq:Etilde_SV_FWHQET}
\end{align}
For $S+P$ or $P+V$ condensate, we can obtain the same dispersion relations. 
Compared with the dispersion relations in the conventional HQET, (\ref{eq:Epm_SV_HQET}) and (\ref{eq:Etilde_SV_HQET}), the particle modes $E_\pm(p) $ are the same.
In the FW-HQET, the antiparticle modes are mixed with the heavy particles, and the decoupled modes like Eq.~(\ref{eq:Etilde_SV_HQET}) do not appear.
In the right panel of Fig.~\ref{fig:disp}, we plot an example of the dispersion relations.
Thus, the appearance of the Dirac-cone-like crossing point between $E_-(p)$ and $\tilde{E}_-(p)$ is a unique characteristic in the relativistic Kondo effect based on the FW-HQET, which is different from that in the conventional HQET with the $S+V$ condensates as shown in the previous subsection~\ref{Sec:2-1}.

For the single $S$, $P$, $V$, or $A$ condensate, we get
\begin{align}
E_\pm(p) &\equiv  \frac{1}{2} \left( p + \lambda -\mu \pm \sqrt{\left(p-\lambda-\mu \right)^2 + 4 |\Delta|^2 }\right) \nonumber\\
&\hspace{145pt} (\mathrm{Double}), \\
\tilde{E}_\pm(p) &\equiv \frac{1}{2} \left( -p + \lambda -\mu \pm \sqrt{\left(p+\lambda+\mu \right)^2 + 4 |\Delta|^2 }\right) \nonumber\\
&\hspace{145pt} (\mathrm{Double}).
\end{align}
These agree with those obtained in Refs.~\cite{Feng:2016,Lai:2018}.
Thus, a single condensate leads to the factor $4$ in front of $|\Delta|^2$.

For the $S+P+V$ condensate with $\Delta \equiv \Delta_S^+ = \Delta_S^- = \Delta_P^+ = \Delta_P^- = \Delta_V^+ = \Delta_V^-$, we get
\begin{align}
E_\pm(p) &\equiv  \frac{1}{2} \left( p + \lambda -\mu \pm \sqrt{\left(p-\lambda-\mu \right)^2 + 12 |\Delta|^2 }\right) \nonumber\\
&\hspace{145pt} (\mathrm{Double}), \\
\tilde{E}_\pm(p) &\equiv \frac{1}{2} \left( -p + \lambda -\mu \pm \sqrt{\left(p+\lambda+\mu \right)^2 + 12 |\Delta|^2 }\right) \nonumber\\
&\hspace{145pt} (\mathrm{Double}).
\end{align}
The factor of $3$ among $12$ is caused by the linear combination of the three types of condensates.

We comment on the thermodynamic potential for each condensate. 
For the ``non-combinatorial" condensates such as $S$, $P$, $V$ and $A$, the condensate energy part of the thermodynamic potential is $2|\Delta|^2/G$ (the factor of $2$ comes from $\Delta^+$ and $\Delta^-$).
The two-combinatorial ($S+P$, $S+V$, and $P+V$) and three-combinatorial ($S+P+V$) condensates have $4|\Delta|^2/G$ and $6|\Delta|^2/G$, respectively.
Therefore, the coefficient in front of $|\Delta|^2$ is scaled as the number of the condensates, so that $S$, $P$, $V$, $A$, $S+P$, $S+V$, $P+V$ and $S+P+V$ are thermodynamically equivalent to each other.

For the $V+A$ condensate with $\Delta \equiv \Delta_V^+ = \Delta_V^- = \Delta_A^+ = \Delta_A^-$, we get
\begin{align}
E_\pm (p) &\equiv \frac{1}{2} \left( p + \lambda -\mu \pm \sqrt{\left(p-\lambda-\mu \right)^2 + 16 |\Delta|^2 }\right) \nonumber\\
&\hspace{145pt} (\mathrm{Single}), \label{eq:Epm_VA_FWHQET}\\
\tilde{E}_\pm (p) &\equiv \frac{1}{2} \left( - p + \lambda -\mu \pm \sqrt{\left(p+\lambda+\mu \right)^2 + 16 |\Delta|^2 }\right) \nonumber\\
&\hspace{145pt} (\mathrm{Single}), \label{eq:Epmtilde_VA_FWHQET}\\
E(p)     &\equiv p-\mu \ \ \ (\mathrm{Single}), \label{eq:E_VA_FWHQET} \\
\tilde{E}(p)     &\equiv -p-\mu \ \ \ (\mathrm{Single}), \label{eq:Etilde_VA_FWHQET} \\
E_Q  &\equiv \lambda \ \ \ (\mathrm{Double}). \label{eq:EQ_VA_FWHQET}
\end{align}
Thus, we find the four mixing modes and four decoupled modes.
For the mixing modes, the factor of $16$ in front of $|\Delta|^2$ might be characteristic.
Each mode except for $E_Q$ does not degenerate, and both the particle $E(p)$ and antiparticle $\tilde{E}(p)$ modes are decoupled from the condensate.

\section{Discussion} \label{Sec:3}

\subsection{Possible physical situations for two HQETs}
In Sec.~\ref{Sec:2}, we have discussed the relativistic Kondo effects based on the two different types of HQETs, i.e., the conventional HQET and the FW-HQET.
The former includes only the particle component, while the latter includes both the particle and antiparticle components.
In the former, the heavy-particle number density is controlled by the chemical potential $\lambda$ (measured in the residual momentum space).
Similarly, in the latter, the heavy-particle number density and the heavy-antiparticle number density are controlled by $\lambda_{+}$ and $\lambda_{-}$, respectively.
Thus, the FW-HQET can describe environments where there exist not only heavy-particle impurities but also the heavy-antiparticle impurities.

In other words, we can consider different physical situations by utilizing the difference between the two types of HQETs.
For example, the FW-HQET with nonzero $\lambda_{+}$ and $\lambda_{-}$ can allow coexistence of both heavy particles and heavy antiparticles, and then both the particle density and antiparticle density can be nonzero at the same time.
On the other hand, in the conventional HQET, such heavy-antiparticle density is always absent by the definition of the heavy field, and only the heavy-particle density can be nonzero.
Thus, the physical situation described by the FW-HQET can be different from that in the conventional HQET.

\subsection{Possible forms of Kondo condensates}
The possible condensates that can induce simple dispersion relations are summarized in Table \ref{Tab_list}.
Here, we compare the three situations: (i) mixing between a light Dirac fermion and a heavy fermion in the conventional HQET, (ii) mixing between a light Dirac fermion and a heavy fermion in the FW-HQET, and (iii) mixing between two light Dirac fermions.
For the case (iii), we can get a simple form of dispersion relations, but the possible forms are different from the typical dispersion relations in the Kondo effects (see the Appendix for a detailed discussion).

\begin{table}[t!]
\centering
\caption{The list of possible particle-antiparticle condensates in three cases: (i) mixing between a light Dirac fermion and a heavy fermion defined in the conventional heavy-quark effective theory (con.HQET), (ii) mixing between a light Dirac fermion and a heavy fermion in the heavy-quark effective theory with the FW transformation (FW-HQET), and (iii) mixing between two light Dirac fermions (2-Dirac).
$S$, $P$, $V$, and $A$ denote the scalar ($1$), pseudoscalar ($i \gamma_5$), vector-hedgehog ($\hat{p} \cdot \vec{\gamma}$), and axial-vector--hedgehog ($\hat{p} \cdot \vec{\gamma} \gamma_5$) condensates, respectively.
For ``$\checkmark$", we can get a simple form of the dispersion relations.
For ``$-$", we cannot get simple dispersion relations.}
\begin{tabular}{cccc}
\hline\hline
Condensate & con.HQET & FW-HQET & 2-Dirac \\
\hline
$S$  & $-$  & $\checkmark$ & $\checkmark$ \\
$P$  & $-$  & $\checkmark$ & $\checkmark$\\
$V$  & $-$  & $\checkmark$ & $\checkmark$ \\
$A$  & $-$  & $\checkmark$ & $\checkmark$ \\
\hline
$S+P$  & $-$ & $\checkmark$ & $\checkmark$ \\
$S+V$  & $\checkmark$  & $\checkmark$ & $\checkmark$ \\
$S+A$  & $-$  & $-$     & $\checkmark$ \\
$P+V$  & $-$  & $\checkmark$ & $\checkmark$\\
$P+A$  & $-$  & $-$     & $\checkmark$\\
$V+A$  & $\checkmark$ & $\checkmark$ & $\checkmark$ \\
\hline
$S+P+V$  & $-$ & $\checkmark$ & $\checkmark$ \\
$S+P+A$  & $-$ & $-$     & $\checkmark$ \\
$S+V+A$  & $-$ & $-$     & $-$ \\
$P+V+A$  & $-$ & $-$     & $-$ \\
\hline
$S+P+V+A$  & $-$ & $-$ & $-$ \\
\hline\hline
\end{tabular}
\label{Tab_list}
\end{table}

As shown in Table \ref{Tab_list}, in the conventional HQET, only for the $S + V$ and $V + A$ condensates, we can obtain a simple form of dispersion relations of the quasiparticles, where $S + V$ corresponds to the {\it Ansatz} found in Ref.~\cite{Yasui:2016svc,Yasui:2017izi}.
On the other hand, $V + A$ is newly found in this work.
If we assume the parity symmetry of the ground state (in other words, $\Delta$ is even parity), only the $S + V$ condensate is allowed.\footnote{From the definition of vector-hedgehog condensate, $G \langle \bar{\psi}_1 \vec{\gamma} \Psi_v^+  \rangle \equiv \Delta_V \hat{p}$, the condensate $\langle \bar{\psi}_1 \vec{\gamma} \Psi_v^+  \rangle$ is odd parity, and the gap $\Delta_V$ is even parity.}
Thus, in the conventional HQET, only the hedgehog-type condensates are a promising candidate as a ground state of the Kondo effect. 
This limitation is because the heavy-fermion field is formally a two-component spinor due to the particle (or antiparticle) projection.

We emphasize that, in the conventional HQET, for non-combinatorial condensates such as $S$, $P$, $V$, and $A$, we cannot get simple dispersions.
An interpretation is as follows.
When we first consider the vector hedgehog, the vector hedgehog mixes the light lower component and heavy particle, but it cannot mix the light upper component and heavy antiparticle due to the particle projection of the heavy fermion.\footnote{The ``upper" and ``lower" mean the upper and lower two components of the four-component Dirac spinor.
}
To compensate the mixing lost for the light upper component, the mixing by the $S$ or $A$ condensate is required ($P$ condensate does not mix them).
Thus, in order to describe a ground state in this limit by using condensates, we need combinatorial condensates as a linear combination of ($S$, $P$, $V$, and $A$).

In the FW-HQET, the heavy field consists of the four-component Dirac spinor as usual.
As a result, we can get simple forms of dispersion relations for non-combinatorial condensates such as $S$, $P$, $V$, and $A$ as the ground state in the Kondo effects.
Also, other linear combinations such as $S+P$, $P+V$, and $S+P+V$ become possible.
If we assume the parity symmetry of the ground state, only the $S$, $V$, and $S + V$ condensates are allowed.
These ground states ($S$, $V$, and $S + V$) are equivalent (or degenerate) in the thermodynamic potential.

\subsection{Relation to topological Kondo insulators}
Finally, we discuss the relation between the relativistic Kondo condensates and hybridizations in nonrelativistic TKIs~\cite{DzeroSunGalitskiColeman2010,DzeroSunColemanGalitski2012}.
The TKIs are induced by a band inversion between even-parity $d$-electron bands and odd-parity $f$-electron bands, which is driven by a spin-orbit interaction stronger than the typical gap of Kondo insulators.

\begin{figure}[tb!]
    \center
    \begin{minipage}[t]{1.0\columnwidth}
        \begin{center}
            \includegraphics[clip, width=1.0\columnwidth]{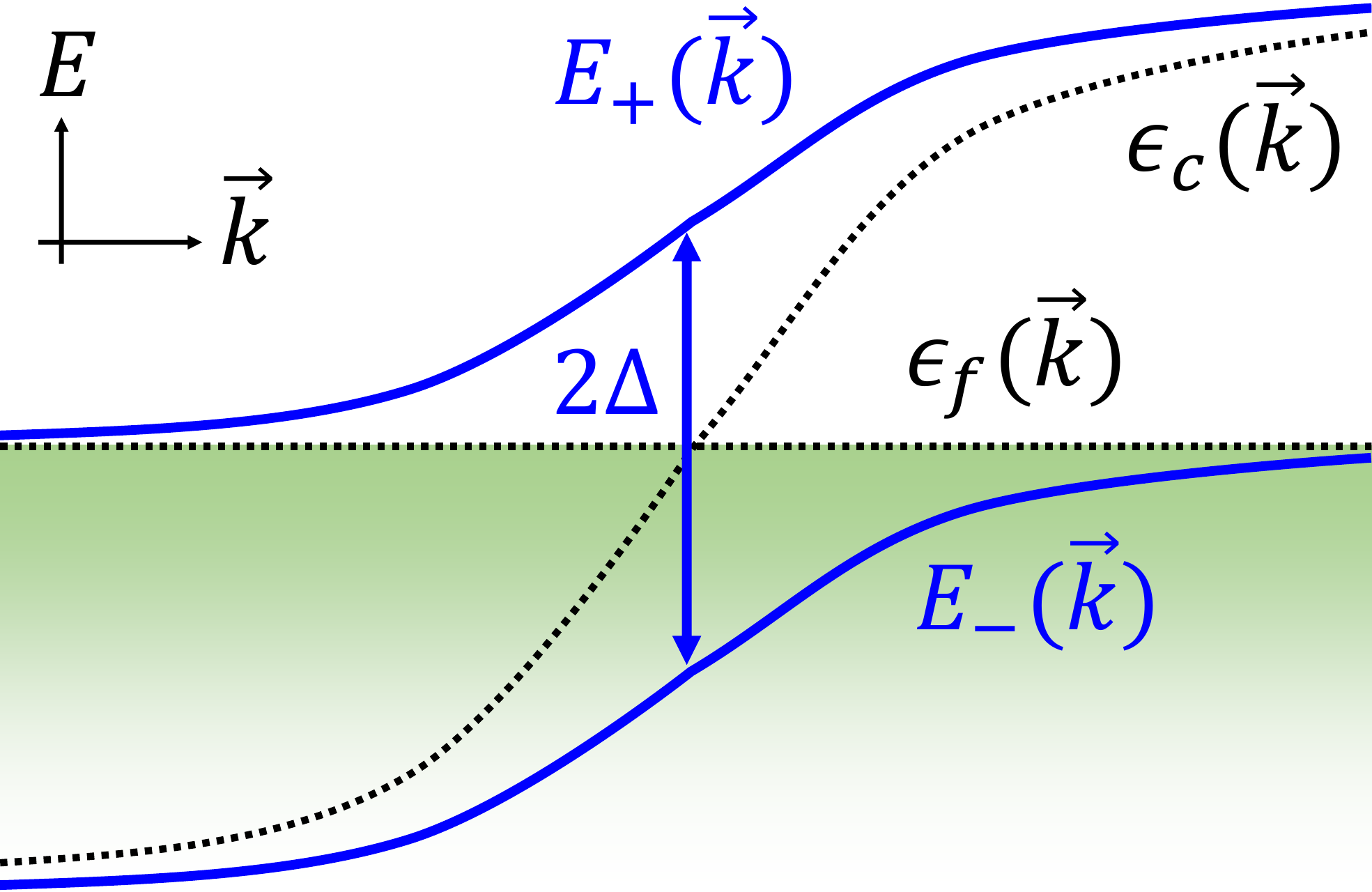}
        \end{center}
    \end{minipage}%
    \caption{Schematic picture of dispersion relations in nonrelativistic topological Kondo insulators.
The $d$-electron band $\epsilon_c(\vec{k})$ and $f$-electron band $\epsilon_f(\vec{k})$ are hybridized, and the quasiparticle bands $E_\pm(\vec{k})$ with a gap $\Delta$ appear.
The shaded region means the Fermi sea.}
\label{fig:TKI}
\end{figure}

We consider a minimal Hamiltonian of the TKI \cite{DzeroXiaGalitskiColeman2016}, where conduction $d$ electrons are hybridized with $f$ electrons:
\begin{align}
& H_\mathrm{TKI} \equiv
\sum_{\vec{k}} \left(
\begin{array}{c}
c_d^\dagger (\vec{k}) \\
c_f^\dagger (\vec{k}) \\
\end{array}
\right)^T \Gamma (\vec{k}) \left(
\begin{array}{c}
c_d (\vec{k}) \\
c_f (\vec{k}) \\
\end{array}
\right) \nonumber\\
& \equiv
\sum_{\vec{k}} \left(
\begin{array}{cc}
c_d^\dagger(\vec{k}) \\
c_f^\dagger(\vec{k}) \\
\end{array}
\right)^T \left(
\begin{array}{cc}
\epsilon_c(\vec{k})  & \Delta \vec{d}(\vec{k}) \cdot \vec{\sigma} \\
\Delta \vec{d}(\vec{k}) \cdot \vec{\sigma} & \epsilon_f(\vec{k}) \\
\end{array}
\right) \left(
\begin{array}{cc}
c_d(\vec{k}) \\
c_f(\vec{k}) \\
\end{array}
\right), \label{eq:TKI}
\end{align}
where each of the $d$ and $f$ electrons consist of two components (or a Kramers doublet) in pseudospin space: $(c_d^\dagger,c_f^\dagger) \equiv (c^\dagger_{d\uparrow},c^\dagger_{d\downarrow},c^\dagger_{f\uparrow},c^\dagger_{f\downarrow})$.
The energy levels, $\epsilon_c(\vec{k})$ and $ \epsilon_f(\vec{k})$, and the hybridization $\Delta \vec{d}(\vec{k}) \cdot \vec{\sigma}$ with a ``form factor" $\vec{d}(\vec{k}) \cdot \vec{\sigma}$ and a strength $\Delta$ are $2\times2$ matrices in pseudospin space.
This hybridization is induced by the (spin) exchange interaction between the $d$ electron and $f$ electron. 
For the cubic lattice, the three-momentum vector is $\vec{d}(\vec{k}) = (\sin k_x,\sin k_y,\sin k_z)$, but it is approximated to be $\vec{d}(\vec{k}) \sim (k_x,k_y,k_z)$ at a small $\vec{k}$.
Note that if the Fermi level of quasiparticles is inside the hybridization gap, the ground state of the Hamiltonian (\ref{eq:TKI}) is a Kondo insulator with the hedgehog-type condensate in momentum space.
From this Hamiltonian, the dispersion relations for the quasiparticles are
\begin{align}
& E_\pm(\vec{k}) \equiv \nonumber\\
&\frac{1}{2} \left(\epsilon_c(\vec{k}) + \epsilon_f(\vec{k}) \pm \sqrt{\left(\epsilon_c(\vec{k}) - \epsilon_f(\vec{k})  \right)^2 + 4 \Delta^2 |\vec{d}(\vec{k})|^2}\right) \nonumber\\
&\hspace{175pt} (\mathrm{Double}).
\end{align}
In Fig.~\ref{fig:TKI}, we show a schematic picture for the band structures before and after the hybridization.

Next, we compare the Hamiltonian (\ref{eq:TKI}) for nonrelativistic TKIs and the Lagrangians for relativistic Kondo effects.
For the inverse propagator~(\ref{eq:Ginverse_SPVA_HQET}) from the conventional HQET, the corresponding Hamiltonian form is
\begin{widetext}
\begin{align}
\Gamma_\mathrm{HQET}^{S,P,V,A} (\vec{p})
\equiv 
\left(
\begin{array}{ccc}
-\mu                        & \vec{p}\cdot \vec{\sigma}& -\Delta_{S}^{\ast} + \Delta_A^{\ast} \hat{p}\cdot \vec{\sigma}  \\
\vec{p}\cdot \vec{\sigma} & -\mu & -\Delta_V^{\ast} \hat{p}\cdot \vec{\sigma} +i\Delta^{\ast}_P \\
-\Delta_S +\Delta_A \hat{p}\cdot \vec{\sigma}  & -\Delta_V \hat{p}\cdot \vec{\sigma} -i\Delta_P &  \lambda \\
\end{array}
\right)
.
\label{eq:Hamiltonian_SPVA_HQET}
\end{align}
Note that this form is a $6\times6$ matrix, and it consists of the spin-doublets for the light-particle, light-antiparticle, and heavy-particle components.
For the inverse propagator~(\ref{eq:Ginverse_SPVA_FW}) from the FW-HQET, the corresponding Hamiltonian form is
\begin{eqnarray}
 \Gamma_\mathrm{FW}^{S,P,V,A} (\vec{p})
\equiv
 \left(
\begin{array}{cccc}
-\mu                        & \vec{p}\cdot \vec{\sigma}& -\Delta_{S}^{+\ast} + \Delta_A^{+\ast} \hat{p}\cdot \vec{\sigma} & -\Delta_V^{-\ast} \hat{p}\cdot \vec{\sigma} -i\Delta^{-\ast}_P \\
\vec{p}\cdot \vec{\sigma} & -\mu & -\Delta_V^{+\ast} \hat{p}\cdot \vec{\sigma} +i\Delta^{+\ast}_P & \Delta_S^{-\ast} + \Delta_A^{-\ast} \hat{p}\cdot \vec{\sigma} \\
-\Delta_S^+ +\Delta_A^+ \hat{p}\cdot \vec{\sigma}  & -\Delta_V^+ \hat{p}\cdot \vec{\sigma} -i\Delta_P^+ &  \lambda_+        & 0 \\
-\Delta_V^- \hat{p}\cdot \vec{\sigma}+i\Delta_P^- & \Delta_S^- +\Delta_A^- \hat{p}\cdot \vec{\sigma} & 0  & \lambda_- \\
\end{array}
\right)
. \label{eq:Hamiltonian_SPVA_FW}
\end{eqnarray}
\end{widetext}
Note that this form is a $8\times8$ matrix, and it consists of the spin-doublets for the light-particle, light-antiparticle, heavy-particle, and heavy-antiparticle components.
Thus, in the relativistic light Dirac fermions, by definition, the spin-momentum locking, $\vec{p}\cdot \vec{\sigma}$, appears in the $(1,2)$ and $(2,1)$ components in the matrix representation.
In addition, when a relativistic Kondo effect is realized, we can consider hedgehog-type hybridization as a Kondo condensate, which is a similar situation to the hybridization in TKIs.

We emphasize that the mechanism realizing the hedgehog condensates in momentum space for nonrelativistic TKIs and relativistic Kondo effects is almost the same.
In the TKI Hamiltonian (\ref{eq:TKI}), the odd-parity condensate, $\Delta \vec{d}(\vec{k}) \cdot \vec{\sigma}$, is induced by the pairing between the even-parity $d$ electron and odd-parity $f$ electron, as long as the spatial inversion symmetry of the whole Hamiltonian is preserved.
On the other hand, the Hamiltonians for the relativistic Kondo effects, (\ref{eq:Hamiltonian_SPVA_HQET}) and (\ref{eq:Hamiltonian_SPVA_FW}), include the hedgehog condensates such as $\Delta_V^{(\ast)} \hat{p}\cdot \vec{\sigma}$ and $\Delta_A^{(\ast)} \hat{p}\cdot \vec{\sigma}$.
The vector hedgehog $\Delta_V^{(\ast)} \hat{p}\cdot \vec{\sigma}$ and pseudoscalar $\Delta_P^{(\ast)}$ condensates with odd parity are composed of the light lower component with a parity and the heavy-particle component with the opposite parity (or light upper and heavy-antiparticle components) since the parity of the upper component is opposite to that of the lower component.
The scalar $\Delta_S^{(\ast)}$ and axial-vector--hedgehog $\Delta_A^{(\ast)} \hat{p}\cdot \vec{\sigma}$ condensates with even parity, where $\Delta_A^{(\ast)}$ is odd parity, are composed of the light upper and heavy-particle (or the light lower and heavy-antiparticle) components.
When the spatial inversion symmetry of the ground state is preserved (namely, $\Delta$ is even parity), only the scalar and vector hedgehog condensates are possible.
Thus, the Hamiltonian for the vector hedgehog condensate is similar to that for the TKI.

\section{Conclusion and outlook} \label{Sec:4}

In this paper, we showed that the two types of relativistic Kondo effects can be realized based on the two types of HQETs, as shown in Fig.~\ref{fig:disp}, where ``relativistic" means that the light fermions are treated as the Dirac field, and the heavy-impurity fields are in the nonrelativistic limits of the massive Dirac fermion.
The essential difference between the two types of HQETs is the existence or absence of the heavy antiparticle component.
The two Kondo effects are similar to each other near the Fermi surface but slightly differ in the structure at low momentum.

In the conventional HQET, we showed that the combinatorial hedgehog-type condensates such as the $S+V$ condensate~\cite{Yasui:2016svc,Yasui:2017izi} may be preferentially induced due to the assumption of the particle projection.
On the other hand, in the FW-HQET, the various condensates such as the $S$~\cite{Feng:2016,Lai:2018}, $V$, and $S+V$ condensates can appear.
Although the final conclusion depends on the specific interactions in the focused system, the various Kondo condensate structures studied in this paper will give a guide to investigate the true ground state for a specific system.

One of the open questions is which Kondo effect is closer to the relativistic Kondo effect with the original massive (and sufficiently heavy) Dirac field, such as the QCD Kondo effect with charm or bottom quarks~\cite{Yasui:2013xr,Hattori:2015hka}.
In this sense, a most useful approach is lattice simulations with relativistic heavy Dirac fermions, such as lattice QED or QCD simulations, where we can directly measure the vacuum expectation values of operators composed of a light fermion and a heavy fermion, such as $\langle \bar{\psi}_1 \psi_2 \rangle$ and $\langle \bar{\psi}_1 \vec{\gamma} \psi_2 \rangle$.
On the other hand, in lattice HQET \cite{Eichten:1987xu} simulations, one can apply the particle projection operator to lattice heavy-fermion actions, so that such simulations will be useful for examining nonperturbative properties of the Kondo effect with only particle components of heavy fermions as shown in the subsection~\ref{Sec:2-1}.
Lattice simulations with the projection of the FW-HQET will be also interesting.

The two Kondo effects will be studied by utilizing ultracold-atom simulations.
For studies of nonrelativistic Kondo effects with ultracold atoms, see, e.g., Refs.~\cite{Paredes:2003,Duan:2004,Gorshkov:2010,Bauer:2013,Nishida:2013,Nakagawa:2016,Nakagawa:2018}.
Ultracold-atom simulations implementing relativistic fermions will give us deeper understanding of the relativistic Kondo effects.

In the context of solid-state systems, our discussion can be applied to materials such as Dirac/Weyl semimetals with dilutely doped heavy impurities or superlattice systems with periodic intercalation of heavy-electron layers.
The latter may be designed with material growth techniques such as molecular-beam epitaxy (MBE), building up the crystal layer by layer.
Our findings in this paper imply that parity of heavy impurities, which is determined by the atomic orbital of impurity electrons participating in the Kondo effect, restricts the possible hybridization structure of the Kondo condensate.
We have demonstrated that the resultant hybridization structure affects the band structure.
In particular, the Kondo effect may change the structure of the band crossing, even when the crossing point is located at an energy away from the Fermi level.
The presence or absence of the Dirac point beneath the Fermi level does not significantly affect the thermodynamic properties (such as specific heat) or the stationary transport properties (such as the electric conductivity for direct currents), as long as the Dirac point is far enough away from the Fermi level.
Nevertheless, it can be captured experimentally if an electron living near the Dirac point is excited above the Fermi surface {\it by a finite-frequency external field} such as a laser.
For example, the band structure below the Fermi level may be directly measured by angle-resolved photoemission spectroscopy (ARPES).
Furthermore, electromagnetic responses to the light, such as the optical conductivity and absorption spectrum, also provide information on the structure of the band eigenstates away from the Fermi level.

Once the hybridization structure from the Kondo effect is determined from such measurements, our findings will be helpful in identifying the heavy degrees of freedom participating in the Kondo effect.

\section*{Acknowledgments}
Y.A. is supported by the Leading Initiative for Excellent Young Researchers (LEADER).
S.Y. is supported by the Interdisciplinary Theoretical and Mathematical Sciences Program (iTHEMS) at RIKEN.
This work was supported by Japan Society for the Promotion of Science (JSPS) KAKENHI (Grants No. JP17K05435, No. JP17K14277, and No. JP20K14476).

\appendix
\renewcommand{\theequation}{A\arabic{equation}}
\setcounter{equation}{0}

\begin{widetext}
\section*{Appendix: Mixing between two-flavor Dirac fermions} \label{App:1}
In this Appendix, we study mixing between {\it massless} two-flavor Dirac fermions.
We start from a Lagrangian including two different Dirac fields, $\psi_1$ with chemical potential $\mu$ and $\psi_2$ with $-\lambda$:
\begin{align}
{\cal L}_\mathrm{eff}
\equiv \bar{\psi}_1 \bigl(i\slashed{\partial} +\mu\gamma_0\bigr)\psi_1 + \bar{\psi}_2 \bigl(i\slashed{\partial} - \lambda \gamma_0\bigr)\psi_2 
+G\Bigl[|\bar{\psi}_1\psi_2|^2 +|\bar{\psi}_1 i\gamma_5 \psi_2|^2 + |\bar{\psi}_1 \vec{\gamma}\psi_2|^2  +|\bar{\psi}_1 \vec{\gamma} \gamma_5 \psi|^2\Bigr], \ 
\label{NJL}
\end{align}
where $G$ is the coupling constant.
Here, for the vector and axial-vector interactions, we assumed a situation that the temporal component is suppressed for any reason, and only the spatial components survive, which is a similar situation to the heavy-fermion limit.

Here, we assume the following mean fields for the four-point interaction terms:
\begin{eqnarray}
G \langle \bar{\psi}_1 \psi_2 \rangle  &\equiv& \Delta_S, \\
G \langle \bar{\psi}_1 i\gamma_5 \psi_2 \rangle  &\equiv& \Delta_P, \\
G \langle \bar{\psi}_1 \vec{\gamma} \psi_2  \rangle  &\equiv& \Delta_V \hat{p}, \\
G \langle \bar{\psi}_1 \gamma _5 \vec{\gamma} \psi_2 \rangle  &\equiv& \Delta_A \hat{p},
\end{eqnarray}
where $\hat{p} \equiv \vec{p}/p$ ($p \equiv |\vec{p}|$) is the unit vector of the three-dimensional momentum $\vec{p}$.
The gaps $\Delta_S$, $\Delta_P$, $\Delta_V$, and $\Delta_A$ are a complex number.
We call the third (vector) and fourth (axial-vector) condensates with $\hat{p}$ the {\it hedgehog condensate}.
After assuming the hedgehog {\it Ans\"atze}, $\hat{p}$ is combined with another $\vec{\gamma}$ in the vector four-point interaction, and then we get the factor $\hat{p}\cdot \vec{\gamma}$.
As a result, the mean-field Lagrangian is
\begin{align}
{\cal L}_\mathrm{MF} 
\equiv& \bar{\psi}_1 \bigl( \slashed{p} +\mu\gamma_0\bigr)\psi_1 + \bar{\psi}_2 \bigl(\slashed{p} - \lambda \gamma_0\bigr)\psi_2 
+  \bar{\psi}_2 (\Delta_S + i \Delta_P \gamma_5 + \Delta_V \hat{p}\cdot \vec{\gamma} + \Delta_A \hat{p}\cdot \vec{\gamma} \gamma_5) \psi_1   \nonumber \\
& +  \bar{\psi}_1 (\Delta^\ast_S + i \Delta^\ast_P \gamma_5 + \Delta^\ast_V \hat{p}\cdot \vec{\gamma} + \Delta^\ast_A \hat{p}\cdot \vec{\gamma} \gamma_5) \psi_2
- \frac{1}{G} \left( |\Delta_S|^2 + |\Delta_P|^2 + |\Delta_V|^2 + |\Delta_A|^2 \right). 
\label{2-flavorMF}
\end{align}
From the mean-field Lagrangian (\ref{2-flavorMF}), the inverse propagator of quasiparticles with all the $S$, $P$, $V$, and $A$ condensates is written as
\begin{eqnarray}
 G_\mathrm{MF}^{S,P,V,A}(p_{0},\vec{p}\,)^{-1}
\equiv
 \left(
\begin{array}{cccc}
p_0 + \mu                        & -\vec{p}\cdot \vec{\sigma}& \Delta_S^\ast - \Delta_A^\ast \hat{p}\cdot \vec{\sigma} & \Delta_V^\ast \hat{p}\cdot \vec{\sigma} +i\Delta^\ast_P \\
\vec{p}\cdot \vec{\sigma} & -(p_0 + \mu) & -\Delta_V^\ast \hat{p}\cdot \vec{\sigma} +i\Delta^\ast_P & \Delta_S^\ast + \Delta_A^\ast \hat{p}\cdot \vec{\sigma} \\
\Delta_S -\Delta_A \hat{p}\cdot \vec{\sigma}  & \Delta_V \hat{p}\cdot \vec{\sigma} +i\Delta_P & p_0 - \lambda        & - \vec{p}\cdot \vec{\sigma}  \\
-\Delta_V \hat{p}\cdot \vec{\sigma}+i\Delta_P & \Delta_S +\Delta_A \hat{p}\cdot \vec{\sigma} & \vec{p}\cdot \vec{\sigma}   & -(p_0 - \lambda) \\
\end{array}
\right).
\label{eq:Ginverse_SPVA_2f}
\end{eqnarray}
This propagator consists of $8 \times 8$ components: the two four-component Dirac spinors for $\psi_1$ and $\psi_2$.

From now on, we assume $\Delta \equiv \Delta_S = \Delta_P = \Delta_V = \Delta_A$ for clarity.
From Eq.~(\ref{eq:Ginverse_SPVA_2f}), by solving $\det[G(p_{0})^{-1}]=0$, we get the dispersion relations of quasiparticles.
For example, for the $S+V$ condensate,
\begin{eqnarray}
E_\pm(p) &\equiv&  \frac{1}{2} \left( \lambda -\mu + \sqrt{ 4p^2 +\left(\lambda+\mu \right)^2 +  8 |\Delta|^2 \pm 4 p \sqrt{ \left(\lambda+\mu \right)^2  + 4|\Delta|^2 }}\right) \ \ \ (\mathrm{Double}),
\label{eq:Epm_SV_2flavor} \\
\tilde{E}_\pm(p) &\equiv& \frac{1}{2} \left(  \lambda -\mu - \sqrt{4p^2 + \left(\lambda+\mu \right)^2  + 8 |\Delta|^2 \pm 4 p \sqrt{ \left(\lambda+\mu \right)^2 + 4|\Delta|^2} }\right) \ \ \ (\mathrm{Double}).
\label{eq:Epmtilde_SV_2flavor}
\end{eqnarray}
For the $P+V$ condensate, we can get the same dispersions.
Thus, for mixing between massless Dirac fermions, even when we assume the $S+V$ condensate, we cannot obtain the dispersion relations of the Kondo effect like Eq.~(\ref{eq:Epm_SV_HQET}) in the conventional HQET or Eqs.~(\ref{eq:Epm_SV_FWHQET}) and (\ref{eq:Etilde_SV_FWHQET}) in the FW-HQET.
In other words, the Kondo-type dispersions are induced by the nonrelativistic properties of heavy fermions.

For the single $S$ or $P$ condensate,
\begin{eqnarray}
E_\pm(p) &\equiv&  \frac{1}{2} \left( \lambda -\mu + \sqrt{ \left( \pm 2p + \lambda + \mu \right)^2 +  4 |\Delta|^2 }\right) \ \ \ (\mathrm{Double}),  \label{eq:Epm_S_2flavor} \\
\tilde{E}_\pm(p) &\equiv& \frac{1}{2} \left( \lambda -\mu - \sqrt{ \left( \pm 2p + \lambda + \mu \right)^2 +  4 |\Delta|^2 }\right) \ \ \ (\mathrm{Double}). \label{eq:Epmtilde_S_2flavor}
\end{eqnarray}
For the $P$ condensate, these dispersion relations are equivalent to those \cite{Toublan:2003tt} of quasiparticles with the pion condensate \cite{Son:2000xc} realized at nonzero isospin chemical potential $\mu_I$, where we can check it by using the replacement of $\mu \to \mu_u$ and $-\lambda \to \mu_d$ ($\mu_u$ and $\mu_d$ are the chemical potentials of the up and down quarks, respectively) and the definition of $\mu_I \equiv \frac{1}{2} (\mu_u-\mu_d) = \frac{1}{2}(\mu+\lambda)$.

For the single $V$ or $A$ condensate,
\begin{eqnarray}
E_\pm(p) &\equiv&  \frac{1}{2} \left(2p + \lambda -\mu \pm \sqrt{ \left( \lambda + \mu \right)^2 +  4 |\Delta|^2 }\right) \ \ \ (\mathrm{Double}), \label{eq:Epm_V_2flavor} \\
\tilde{E}_\pm(p) &\equiv&  \frac{1}{2} \left(-2p + \lambda -\mu \pm \sqrt{ \left( \lambda + \mu \right)^2 +  4 |\Delta|^2 }\right) \ \ \ (\mathrm{Double}). \label{eq:Epmtilde_V_2flavor}
\end{eqnarray}
These dispersion relations mean that the vector or axial-vector hedgehog condensate (composed of two massless Dirac fermions) is decoupled from the momentum $p$ of the Dirac fermions.
Therefore, these condensates modify only the vacuum (or condensation) energy in the thermodynamic potential, and the dispersion relations of Dirac fermions are not modified.
This situation is distinct from the hedgehog condensates composed of a light fermion and a heavy impurity, as discussed in the main text.

For the $S+P$ condensate,
\begin{eqnarray}
E_\pm(p) &\equiv&  \frac{1}{2} \left( \lambda -\mu + \sqrt{ \left( \pm 2p + \lambda + \mu \right)^2 +  8 |\Delta|^2 }\right) \ \ \ (\mathrm{Double}), \\
\tilde{E}_\pm(p) &\equiv& \frac{1}{2} \left( \lambda -\mu - \sqrt{ \left( \pm 2p + \lambda + \mu \right)^2 +  8 |\Delta|^2 }\right) \ \ \ (\mathrm{Double}).
\end{eqnarray}
These are the same as Eqs.~(\ref{eq:Epm_S_2flavor}) and (\ref{eq:Epmtilde_S_2flavor}) except for the factor of $2$ in front of $|\Delta|^2$.

For the $S+P+V$ condensate,
\begin{eqnarray}
E_\pm(p) &\equiv&  \frac{1}{2} \left( \lambda -\mu + \sqrt{ 4p^2 +\left(\lambda+\mu \right)^2 +  12 |\Delta|^2 \pm 4 p \sqrt{ \left(\lambda+\mu \right)^2  + 4|\Delta|^2 }}\right) \ \ \ (\mathrm{Double}),
\\
\tilde{E}_\pm(p) &\equiv& \frac{1}{2} \left(  \lambda -\mu - \sqrt{4p^2 + \left(\lambda+\mu \right)^2  + 12 |\Delta|^2 \pm 4 p \sqrt{ \left(\lambda+\mu \right)^2 + 4|\Delta|^2} }\right) \ \ \ (\mathrm{Double}).
\end{eqnarray}
These are the same as Eqs.~(\ref{eq:Epm_SV_2flavor}) and (\ref{eq:Epmtilde_SV_2flavor}) except for the factor of $12 |\Delta|^2$.

For the $S+A$ or $P+A$ condensate,
\begin{eqnarray}
E_\pm(p) &\equiv&  \frac{1}{2} \left( \lambda -\mu + \sqrt{ 4p^2 +\left(\lambda+\mu \right)^2 +  8 |\Delta|^2 \pm 4 \sqrt{ p^2 [\left(\lambda+\mu \right)^2 +4|\Delta|^2]  + 4|\Delta|^4 }}\right) \ \ \ (\mathrm{Double}),
\label{eq:Epm_SA_2flavor} \\
\tilde{E}_\pm(p) &\equiv& \frac{1}{2} \left( \lambda -\mu - \sqrt{ 4p^2 +\left(\lambda+\mu \right)^2 +  8 |\Delta|^2 \pm 4 \sqrt{ p^2 [\left(\lambda+\mu \right)^2 +4|\Delta|^2]  + 4|\Delta|^4 }}\right) \ \ \ (\mathrm{Double}).
\label{eq:Epmtilde_SA_2flavor}
\end{eqnarray}

For the $S+P+A$ condensates,
\begin{eqnarray}
E_\pm(p) &\equiv&  \frac{1}{2} \left( \lambda -\mu + \sqrt{ 4p^2 +\left(\lambda+\mu \right)^2 +  12 |\Delta|^2 \pm 4 \sqrt{ p^2 [\left(\lambda+\mu \right)^2 +4|\Delta|^2]  + 8|\Delta|^4 }}\right) \ \ \ (\mathrm{Double}),
\\
\tilde{E}_\pm(p) &\equiv& \frac{1}{2} \left( \lambda -\mu - \sqrt{ 4p^2 +\left(\lambda+\mu \right)^2 +  12 |\Delta|^2 \pm 4 \sqrt{ p^2 [\left(\lambda+\mu \right)^2 +4|\Delta|^2]  + 8|\Delta|^4 }}\right) \ \ \ (\mathrm{Double}).
\end{eqnarray}
These are the same as Eqs.~(\ref{eq:Epm_SA_2flavor}) and (\ref{eq:Epmtilde_SA_2flavor}) except for the factor of $12 |\Delta|^2$and $8 |\Delta|^4$.

For the $V+A$ condensate,
\begin{eqnarray}
E_\pm(p) &\equiv&  \frac{1}{2} \left(2p + \lambda -\mu \pm \sqrt{ \left( \lambda + \mu \right)^2 +  16 |\Delta|^2 }\right) \ \ \ (\mathrm{Single}), \\
\tilde{E}_\pm(p) &\equiv&  \frac{1}{2} \left(-2p + \lambda -\mu \pm \sqrt{ \left( \lambda + \mu \right)^2 +  16 |\Delta|^2 }\right) \ \ \ (\mathrm{Single}), \\
E_1(p) &\equiv& p - \mu \ \ \ (\mathrm{Single}), \\
E_2(p) &\equiv& p + \lambda \ \ \ (\mathrm{Single}), \\
\tilde{E}_1(p) &\equiv& -p-\mu \ \ \ (\mathrm{Single}), \\
\tilde{E}_2(p) &\equiv& -p+\lambda \ \ \ (\mathrm{Single}).
\end{eqnarray}
Similarly to Eqs.~(\ref{eq:Epm_V_2flavor}) and (\ref{eq:Epmtilde_V_2flavor}), this condensate is decoupled from the momentum $p$ of Dirac fermions.
\end{widetext}

\bibliography{reference}

\begin{thebibliography}{100}%
\makeatletter
\providecommand \@ifxundefined [1]{%
 \@ifx{#1\undefined}
}%
\providecommand \@ifnum [1]{%
 \ifnum #1\expandafter \@firstoftwo
 \else \expandafter \@secondoftwo
 \fi
}%
\providecommand \@ifx [1]{%
 \ifx #1\expandafter \@firstoftwo
 \else \expandafter \@secondoftwo
 \fi
}%
\providecommand \natexlab [1]{#1}%
\providecommand \enquote  [1]{``#1''}%
\providecommand \bibnamefont  [1]{#1}%
\providecommand \bibfnamefont [1]{#1}%
\providecommand \citenamefont [1]{#1}%
\providecommand \href@noop [0]{\@secondoftwo}%
\providecommand \href [0]{\begingroup \@sanitize@url \@href}%
\providecommand \@href[1]{\@@startlink{#1}\@@href}%
\providecommand \@@href[1]{\endgroup#1\@@endlink}%
\providecommand \@sanitize@url [0]{\catcode `\\12\catcode `\$12\catcode
  `\&12\catcode `\#12\catcode `\^12\catcode `\_12\catcode `\%12\relax}%
\providecommand \@@startlink[1]{}%
\providecommand \@@endlink[0]{}%
\providecommand \url  [0]{\begingroup\@sanitize@url \@url }%
\providecommand \@url [1]{\endgroup\@href {#1}{\urlprefix }}%
\providecommand \urlprefix  [0]{URL }%
\providecommand \Eprint [0]{\href }%
\providecommand \doibase [0]{http://dx.doi.org/}%
\providecommand \selectlanguage [0]{\@gobble}%
\providecommand \bibinfo  [0]{\@secondoftwo}%
\providecommand \bibfield  [0]{\@secondoftwo}%
\providecommand \translation [1]{[#1]}%
\providecommand \BibitemOpen [0]{}%
\providecommand \bibitemStop [0]{}%
\providecommand \bibitemNoStop [0]{.\EOS\space}%
\providecommand \EOS [0]{\spacefactor3000\relax}%
\providecommand \BibitemShut  [1]{\csname bibitem#1\endcsname}%
\let\auto@bib@innerbib\@empty
\bibitem [{\citenamefont {Kondo}(1964)}]{Kondo:1964}%
  \BibitemOpen
  \bibfield  {author} {\bibinfo {author} {\bibfnamefont {J.}~\bibnamefont
  {Kondo}},\ }\bibfield  {title} {\enquote {\bibinfo {title} {{Resistance
  Minimum in Dilute Magnetic Alloys}},}\ }\href
  {https://doi.org/10.1143/PTP.32.37} {\bibfield  {journal} {\bibinfo
  {journal} {Prog. Theor. Phys.}\ }\textbf {\bibinfo {volume} {32}},\ \bibinfo
  {pages} {37--49} (\bibinfo {year} {1964})}\BibitemShut {NoStop}%
\bibitem [{\citenamefont {Hewson}(1993)}]{Hewson}%
  \BibitemOpen
  \bibfield  {author} {\bibinfo {author} {\bibfnamefont {A.~C.}\ \bibnamefont
  {Hewson}},\ }\href {https://doi.org/10.1017/CBO9780511470752} {\emph
  {\bibinfo {title} {{The Kondo Problem to Heavy Fermions}}}}\ (\bibinfo
  {publisher} {Cambridge University Press, Cambridge, UK},\ \bibinfo {year}
  {1993})\BibitemShut {NoStop}%
\bibitem [{\citenamefont {Yosida}(1996)}]{Yosida}%
  \BibitemOpen
  \bibfield  {author} {\bibinfo {author} {\bibfnamefont {K.}~\bibnamefont
  {Yosida}},\ }\href {https://www.springer.com/gp/book/9783540606512#aboutBook}
  {\emph {\bibinfo {title} {Theory of Magnetism}}}\ (\bibinfo  {publisher}
  {Springer-Verlag, Berlin},\ \bibinfo {year} {1996})\BibitemShut {NoStop}%
\bibitem [{\citenamefont {Yamada}(2004)}]{Yamada}%
  \BibitemOpen
  \bibfield  {author} {\bibinfo {author} {\bibfnamefont {K.}~\bibnamefont
  {Yamada}},\ }\href {https://doi.org/10.1017/CBO9780511534904} {\emph
  {\bibinfo {title} {{Electron Correlation in Metals}}}}\ (\bibinfo
  {publisher} {Cambridge University Press, Cambridge, UK},\ \bibinfo {year}
  {2004})\BibitemShut {NoStop}%
\bibitem [{\citenamefont {Coleman}(2015)}]{coleman_2015}%
  \BibitemOpen
  \bibfield  {author} {\bibinfo {author} {\bibfnamefont {Piers}\ \bibnamefont
  {Coleman}},\ }\href {https://doi.org/10.1017/CBO9781139020916} {\emph
  {\bibinfo {title} {Introduction to Many-Body Physics}}}\ (\bibinfo
  {publisher} {Cambridge University Press, Cambridge, UK},\ \bibinfo {year}
  {2015})\BibitemShut {NoStop}%
\bibitem [{\citenamefont {Principi}\ \emph {et~al.}(2015)\citenamefont
  {Principi}, \citenamefont {Vignale},\ and\ \citenamefont
  {Rossi}}]{Principi:2015}%
  \BibitemOpen
  \bibfield  {author} {\bibinfo {author} {\bibfnamefont {A.}~\bibnamefont
  {Principi}}, \bibinfo {author} {\bibfnamefont {G.}~\bibnamefont {Vignale}}, \
  and\ \bibinfo {author} {\bibfnamefont {E.}~\bibnamefont {Rossi}},\ }\bibfield
   {title} {\enquote {\bibinfo {title} {{Kondo effect and non-Fermi-liquid
  behavior in Dirac and Weyl semimetals}},}\ }\href
  {https://doi.org/10.1103/PhysRevB.92.041107} {\bibfield  {journal} {\bibinfo
  {journal} {Phys. Rev. B}\ }\textbf {\bibinfo {volume} {92}},\ \bibinfo
  {pages} {041107} (\bibinfo {year} {2015})},\ \Eprint
  {http://arxiv.org/abs/1410.8532} {arXiv:1410.8532 [cond-mat.mes-hall]}
  \BibitemShut {NoStop}%
\bibitem [{\citenamefont
  {Yanagisawa}(2015{\natexlab{a}})}]{Yanagisawa:2015conf}%
  \BibitemOpen
  \bibfield  {author} {\bibinfo {author} {\bibfnamefont {T.}~\bibnamefont
  {Yanagisawa}},\ }\bibfield  {title} {\enquote {\bibinfo {title} {{Dirac
  fermions and Kondo effect}},}\ }\href
  {https://doi.org/10.1088/1742-6596/603/1/012014} {\bibfield  {journal}
  {\bibinfo  {journal} {J. Phys.: Conference Series}\ }\textbf {\bibinfo
  {volume} {603}},\ \bibinfo {pages} {012014} (\bibinfo {year}
  {2015}{\natexlab{a}})},\ \Eprint {http://arxiv.org/abs/1502.07898}
  {arXiv:1502.07898 [cond-mat.str-el]} \BibitemShut {NoStop}%
\bibitem [{\citenamefont {Yanagisawa}(2015{\natexlab{b}})}]{Yanagisawa:2015}%
  \BibitemOpen
  \bibfield  {author} {\bibinfo {author} {\bibfnamefont {T.}~\bibnamefont
  {Yanagisawa}},\ }\bibfield  {title} {\enquote {\bibinfo {title} {{Kondo
  Effect in Dirac Systems}},}\ }\href {https://doi.org/10.7566/JPSJ.84.074705}
  {\bibfield  {journal} {\bibinfo  {journal} {J. Phys. Soc. Jpn.}\ }\textbf
  {\bibinfo {volume} {84}},\ \bibinfo {pages} {074705} (\bibinfo {year}
  {2015}{\natexlab{b}})},\ \Eprint {http://arxiv.org/abs/1505.05295}
  {arXiv:1505.05295 [cond-mat.str-el]} \BibitemShut {NoStop}%
\bibitem [{\citenamefont {Mitchell}\ and\ \citenamefont
  {Fritz}(2015)}]{Mitchell:2015}%
  \BibitemOpen
  \bibfield  {author} {\bibinfo {author} {\bibfnamefont {Andrew~K.}\
  \bibnamefont {Mitchell}}\ and\ \bibinfo {author} {\bibfnamefont {Lars}\
  \bibnamefont {Fritz}},\ }\bibfield  {title} {\enquote {\bibinfo {title}
  {{Kondo effect in three-dimensional Dirac and Weyl systems}},}\ }\href
  {https://doi.org/10.1103/PhysRevB.92.121109} {\bibfield  {journal} {\bibinfo
  {journal} {Phys. Rev. B}\ }\textbf {\bibinfo {volume} {92}},\ \bibinfo
  {pages} {121109} (\bibinfo {year} {2015})},\ \Eprint
  {http://arxiv.org/abs/1506.05491} {arXiv:1506.05491 [cond-mat.str-el]}
  \BibitemShut {NoStop}%
\bibitem [{\citenamefont {Sun}\ \emph {et~al.}(2015)\citenamefont {Sun},
  \citenamefont {Xu}, \citenamefont {Zhang},\ and\ \citenamefont
  {Zhou}}]{Sun:2015}%
  \BibitemOpen
  \bibfield  {author} {\bibinfo {author} {\bibfnamefont {Jin-Hua}\ \bibnamefont
  {Sun}}, \bibinfo {author} {\bibfnamefont {Dong-Hui}\ \bibnamefont {Xu}},
  \bibinfo {author} {\bibfnamefont {Fu-Chun}\ \bibnamefont {Zhang}}, \ and\
  \bibinfo {author} {\bibfnamefont {Yi}~\bibnamefont {Zhou}},\ }\bibfield
  {title} {\enquote {\bibinfo {title} {{A magnetic impurity in a Weyl
  semimetal}},}\ }\href {https://doi.org/10.1103/PhysRevB.92.195124} {\bibfield
   {journal} {\bibinfo  {journal} {Phys. Rev. B}\ }\textbf {\bibinfo {volume}
  {92}},\ \bibinfo {pages} {195124} (\bibinfo {year} {2015})},\ \Eprint
  {http://arxiv.org/abs/1509.05180} {arXiv:1509.05180 [cond-mat.str-el]}
  \BibitemShut {NoStop}%
\bibitem [{\citenamefont {Feng}\ \emph {et~al.}()\citenamefont {Feng},
  \citenamefont {Zhong}, \citenamefont {Dai},\ and\ \citenamefont
  {Si}}]{Feng:2016}%
  \BibitemOpen
  \bibfield  {author} {\bibinfo {author} {\bibfnamefont {Xiao-Yong}\
  \bibnamefont {Feng}}, \bibinfo {author} {\bibfnamefont {Hanting}\
  \bibnamefont {Zhong}}, \bibinfo {author} {\bibfnamefont {Jianhui}\
  \bibnamefont {Dai}}, \ and\ \bibinfo {author} {\bibfnamefont {Qimiao}\
  \bibnamefont {Si}},\ }\bibfield  {title} {\enquote {\bibinfo {title}
  {{Dirac-Kondo semimetals and topological Kondo insulators in the dilute
  carrier limit}},}\ }\href@noop {} {\ }\Eprint
  {http://arxiv.org/abs/1605.02380} {arXiv:1605.02380 [cond-mat.str-el]}
  \BibitemShut {NoStop}%
\bibitem [{\citenamefont {Kanazawa}\ and\ \citenamefont
  {Uchino}(2016)}]{Kanazawa:2016ihl}%
  \BibitemOpen
  \bibfield  {author} {\bibinfo {author} {\bibfnamefont {Takuya}\ \bibnamefont
  {Kanazawa}}\ and\ \bibinfo {author} {\bibfnamefont {Shun}\ \bibnamefont
  {Uchino}},\ }\bibfield  {title} {\enquote {\bibinfo {title} {{Overscreened
  Kondo effect, (color) superconductivity, and Shiba states in Dirac metals and
  quark matter}},}\ }\href {https://doi.org/10.1103/PhysRevD.94.114005}
  {\bibfield  {journal} {\bibinfo  {journal} {Phys. Rev. D}\ }\textbf {\bibinfo
  {volume} {94}},\ \bibinfo {pages} {114005} (\bibinfo {year} {2016})},\
  \Eprint {http://arxiv.org/abs/1609.00033} {arXiv:1609.00033
  [cond-mat.str-el]} \BibitemShut {NoStop}%
\bibitem [{\citenamefont {Lai}\ \emph {et~al.}(2018)\citenamefont {Lai},
  \citenamefont {Grefe}, \citenamefont {Paschen},\ and\ \citenamefont
  {Si}}]{Lai:2018}%
  \BibitemOpen
  \bibfield  {author} {\bibinfo {author} {\bibfnamefont {Hsin-Hua}\
  \bibnamefont {Lai}}, \bibinfo {author} {\bibfnamefont {Sarah~E.}\
  \bibnamefont {Grefe}}, \bibinfo {author} {\bibfnamefont {Silke}\ \bibnamefont
  {Paschen}}, \ and\ \bibinfo {author} {\bibfnamefont {Qimiao}\ \bibnamefont
  {Si}},\ }\bibfield  {title} {\enquote {\bibinfo {title} {{Weyl-Kondo
  semimetal in heavy-fermion systems}},}\ }\href
  {https://doi.org/10.1073/pnas.1715851115} {\bibfield  {journal} {\bibinfo
  {journal} {Proc. Natl. Acad. Sci.}\ }\textbf {\bibinfo {volume} {115}},\
  \bibinfo {pages} {93--97} (\bibinfo {year} {2018})},\ \Eprint
  {http://arxiv.org/abs/1612.03899} {arXiv:1612.03899 [cond-mat.str-el]}
  \BibitemShut {NoStop}%
\bibitem [{\citenamefont {Ok}\ \emph {et~al.}()\citenamefont {Ok},
  \citenamefont {Legner}, \citenamefont {Neupert},\ and\ \citenamefont
  {Cook}}]{Ok:2017}%
  \BibitemOpen
  \bibfield  {author} {\bibinfo {author} {\bibfnamefont {Seulgi}\ \bibnamefont
  {Ok}}, \bibinfo {author} {\bibfnamefont {Markus}\ \bibnamefont {Legner}},
  \bibinfo {author} {\bibfnamefont {Titus}\ \bibnamefont {Neupert}}, \ and\
  \bibinfo {author} {\bibfnamefont {Ashley~M.}\ \bibnamefont {Cook}},\
  }\bibfield  {title} {\enquote {\bibinfo {title} {{Magnetic Weyl and Dirac
  Kondo semimetal phases in heterostructures}},}\ }\href@noop {} {\ }\Eprint
  {http://arxiv.org/abs/1703.03804} {arXiv:1703.03804 [cond-mat.str-el]}
  \BibitemShut {NoStop}%
\bibitem [{\citenamefont {Ma}\ \emph {et~al.}(2018)\citenamefont {Ma},
  \citenamefont {Chen}, \citenamefont {Liu},\ and\ \citenamefont
  {Xie}}]{PhysRevB.97.045148}%
  \BibitemOpen
  \bibfield  {author} {\bibinfo {author} {\bibfnamefont {Da}~\bibnamefont
  {Ma}}, \bibinfo {author} {\bibfnamefont {Hua}\ \bibnamefont {Chen}}, \bibinfo
  {author} {\bibfnamefont {Haiwen}\ \bibnamefont {Liu}}, \ and\ \bibinfo
  {author} {\bibfnamefont {X.~C.}\ \bibnamefont {Xie}},\ }\bibfield  {title}
  {\enquote {\bibinfo {title} {{Kondo effect with Weyl semimetal Fermi
  arcs}},}\ }\href {https://doi.org/10.1103/PhysRevB.97.045148} {\bibfield
  {journal} {\bibinfo  {journal} {Phys. Rev. B}\ }\textbf {\bibinfo {volume}
  {97}},\ \bibinfo {pages} {045148} (\bibinfo {year} {2018})},\ \Eprint
  {http://arxiv.org/abs/1709.08008} {arXiv:1709.08008 [cond-mat.str-el]}
  \BibitemShut {NoStop}%
\bibitem [{\citenamefont {Li}\ \emph {et~al.}(2018{\natexlab{a}})\citenamefont
  {Li}, \citenamefont {Sun}, \citenamefont {Wang}, \citenamefont {Xu},
  \citenamefont {Luo},\ and\ \citenamefont {Chen}}]{PhysRevB.98.075110}%
  \BibitemOpen
  \bibfield  {author} {\bibinfo {author} {\bibfnamefont {Lin}\ \bibnamefont
  {Li}}, \bibinfo {author} {\bibfnamefont {Jin-Hua}\ \bibnamefont {Sun}},
  \bibinfo {author} {\bibfnamefont {Zhen-Hua}\ \bibnamefont {Wang}}, \bibinfo
  {author} {\bibfnamefont {Dong-Hui}\ \bibnamefont {Xu}}, \bibinfo {author}
  {\bibfnamefont {Hong-Gang}\ \bibnamefont {Luo}}, \ and\ \bibinfo {author}
  {\bibfnamefont {Wei-Qiang}\ \bibnamefont {Chen}},\ }\bibfield  {title}
  {\enquote {\bibinfo {title} {{Magnetic states and Kondo screening in Weyl
  semimetals with chiral anomaly}},}\ }\href
  {https://doi.org/10.1103/PhysRevB.98.075110} {\bibfield  {journal} {\bibinfo
  {journal} {Phys. Rev. B}\ }\textbf {\bibinfo {volume} {98}},\ \bibinfo
  {pages} {075110} (\bibinfo {year} {2018}{\natexlab{a}})}\BibitemShut
  {NoStop}%
\bibitem [{\citenamefont {L\"u}\ \emph {et~al.}(2019)\citenamefont {L\"u},
  \citenamefont {Deng}, \citenamefont {Ke}, \citenamefont {Guo},\ and\
  \citenamefont {Zhang}}]{PhysRevB.99.115109}%
  \BibitemOpen
  \bibfield  {author} {\bibinfo {author} {\bibfnamefont {Hai-Feng}\
  \bibnamefont {L\"u}}, \bibinfo {author} {\bibfnamefont {Ying-Hua}\
  \bibnamefont {Deng}}, \bibinfo {author} {\bibfnamefont {Sha-Sha}\
  \bibnamefont {Ke}}, \bibinfo {author} {\bibfnamefont {Yong}\ \bibnamefont
  {Guo}}, \ and\ \bibinfo {author} {\bibfnamefont {Huai-Wu}\ \bibnamefont
  {Zhang}},\ }\bibfield  {title} {\enquote {\bibinfo {title} {{Quantum impurity
  in topological multi-Weyl semimetals}},}\ }\href
  {https://doi.org/10.1103/PhysRevB.99.115109} {\bibfield  {journal} {\bibinfo
  {journal} {Phys. Rev. B}\ }\textbf {\bibinfo {volume} {99}},\ \bibinfo
  {pages} {115109} (\bibinfo {year} {2019})}\BibitemShut {NoStop}%
\bibitem [{\citenamefont {Kim}\ and\ \citenamefont {Han}(2019)}]{KIM2019236}%
  \BibitemOpen
  \bibfield  {author} {\bibinfo {author} {\bibfnamefont {Ki-Seok}\ \bibnamefont
  {Kim}}\ and\ \bibinfo {author} {\bibfnamefont {Jae-Ho}\ \bibnamefont {Han}},\
  }\bibfield  {title} {\enquote {\bibinfo {title} {{Interplay between chiral
  magnetic and Kondo effects in Weyl metal phase}},}\ }\href
  {https://doi.org/10.1016/j.cap.2018.08.013} {\bibfield  {journal} {\bibinfo
  {journal} {Curr. Appl. Phys.}\ }\textbf {\bibinfo {volume} {19}},\ \bibinfo
  {pages} {236 -- 240} (\bibinfo {year} {2019})}\BibitemShut {NoStop}%
\bibitem [{\citenamefont {Grefe}\ \emph {et~al.}(2020)\citenamefont {Grefe},
  \citenamefont {Lai}, \citenamefont {Paschen},\ and\ \citenamefont
  {Si}}]{Grefe:2019}%
  \BibitemOpen
  \bibfield  {author} {\bibinfo {author} {\bibfnamefont {Sarah~E.}\
  \bibnamefont {Grefe}}, \bibinfo {author} {\bibfnamefont {Hsin-Hua}\
  \bibnamefont {Lai}}, \bibinfo {author} {\bibfnamefont {Silke}\ \bibnamefont
  {Paschen}}, \ and\ \bibinfo {author} {\bibfnamefont {Qimiao}\ \bibnamefont
  {Si}},\ }\bibfield  {title} {\enquote {\bibinfo {title} {{Weyl-Kondo
  semimetals in nonsymmorphic systems}},}\ }\href
  {https://doi.org/10.1103/PhysRevB.101.075138} {\bibfield  {journal} {\bibinfo
   {journal} {Phys. Rev. B}\ }\textbf {\bibinfo {volume} {101}},\ \bibinfo
  {pages} {075138} (\bibinfo {year} {2020})},\ \Eprint
  {http://arxiv.org/abs/1911.01400} {arXiv:1911.01400 [cond-mat.str-el]}
  \BibitemShut {NoStop}%
\bibitem [{\citenamefont {Yasui}\ and\ \citenamefont
  {Sudoh}(2013)}]{Yasui:2013xr}%
  \BibitemOpen
  \bibfield  {author} {\bibinfo {author} {\bibfnamefont {S.}~\bibnamefont
  {Yasui}}\ and\ \bibinfo {author} {\bibfnamefont {K.}~\bibnamefont {Sudoh}},\
  }\bibfield  {title} {\enquote {\bibinfo {title} {{Heavy-quark dynamics for
  charm and bottom flavor on the Fermi surface at zero temperature}},}\ }\href
  {https://doi.org/10.1103/PhysRevC.88.015201} {\bibfield  {journal} {\bibinfo
  {journal} {Phys. Rev. C}\ }\textbf {\bibinfo {volume} {88}},\ \bibinfo
  {pages} {015201} (\bibinfo {year} {2013})},\ \Eprint
  {http://arxiv.org/abs/1301.6830} {arXiv:1301.6830 [hep-ph]} \BibitemShut
  {NoStop}%
\bibitem [{\citenamefont {Yasui}(2016)}]{Yasui:2016ngy}%
  \BibitemOpen
  \bibfield  {author} {\bibinfo {author} {\bibfnamefont {Shigehiro}\
  \bibnamefont {Yasui}},\ }\bibfield  {title} {\enquote {\bibinfo {title}
  {{Kondo effect in charm and bottom nuclei}},}\ }\href
  {https://doi.org/10.1103/PhysRevC.93.065204} {\bibfield  {journal} {\bibinfo
  {journal} {Phys. Rev. C}\ }\textbf {\bibinfo {volume} {93}},\ \bibinfo
  {pages} {065204} (\bibinfo {year} {2016})},\ \Eprint
  {http://arxiv.org/abs/1602.00227} {arXiv:1602.00227 [hep-ph]} \BibitemShut
  {NoStop}%
\bibitem [{\citenamefont {Yasui}\ and\ \citenamefont
  {Sudoh}(2017)}]{Yasui:2016hlz}%
  \BibitemOpen
  \bibfield  {author} {\bibinfo {author} {\bibfnamefont {Shigehiro}\
  \bibnamefont {Yasui}}\ and\ \bibinfo {author} {\bibfnamefont {Kazutaka}\
  \bibnamefont {Sudoh}},\ }\bibfield  {title} {\enquote {\bibinfo {title}
  {{Kondo effect of $\bar{D}_{s}$ and $\bar{D}_{s}^{\ast}$ mesons in nuclear
  matter}},}\ }\href {https://doi.org/10.1103/PhysRevC.95.035204} {\bibfield
  {journal} {\bibinfo  {journal} {Phys. Rev. C}\ }\textbf {\bibinfo {volume}
  {95}},\ \bibinfo {pages} {035204} (\bibinfo {year} {2017})},\ \Eprint
  {http://arxiv.org/abs/1607.07948} {arXiv:1607.07948 [hep-ph]} \BibitemShut
  {NoStop}%
\bibitem [{\citenamefont {Yasui}\ and\ \citenamefont
  {Miyamoto}(2019)}]{Yasui:2019ogk}%
  \BibitemOpen
  \bibfield  {author} {\bibinfo {author} {\bibfnamefont {Shigehiro}\
  \bibnamefont {Yasui}}\ and\ \bibinfo {author} {\bibfnamefont {Tomokazu}\
  \bibnamefont {Miyamoto}},\ }\bibfield  {title} {\enquote {\bibinfo {title}
  {{Spin-isospin Kondo effects for $\Sigma_{c}$ and $\Sigma_{c}^{\ast}$ baryons
  and $\bar{D}$ and $\bar{D}^{\ast}$ mesons}},}\ }\href
  {https://doi.org/10.1103/PhysRevC.100.045201} {\bibfield  {journal} {\bibinfo
   {journal} {Phys. Rev. C}\ }\textbf {\bibinfo {volume} {100}},\ \bibinfo
  {pages} {045201} (\bibinfo {year} {2019})},\ \Eprint
  {http://arxiv.org/abs/1905.02478} {arXiv:1905.02478 [hep-ph]} \BibitemShut
  {NoStop}%
\bibitem [{\citenamefont {Hattori}\ \emph {et~al.}(2015)\citenamefont
  {Hattori}, \citenamefont {Itakura}, \citenamefont {Ozaki},\ and\
  \citenamefont {Yasui}}]{Hattori:2015hka}%
  \BibitemOpen
  \bibfield  {author} {\bibinfo {author} {\bibfnamefont {Koichi}\ \bibnamefont
  {Hattori}}, \bibinfo {author} {\bibfnamefont {Kazunori}\ \bibnamefont
  {Itakura}}, \bibinfo {author} {\bibfnamefont {Sho}\ \bibnamefont {Ozaki}}, \
  and\ \bibinfo {author} {\bibfnamefont {Shigehiro}\ \bibnamefont {Yasui}},\
  }\bibfield  {title} {\enquote {\bibinfo {title} {{QCD Kondo effect: Quark
  matter with heavy-flavor impurities}},}\ }\href
  {https://doi.org/10.1103/PhysRevD.92.065003} {\bibfield  {journal} {\bibinfo
  {journal} {Phys. Rev. D}\ }\textbf {\bibinfo {volume} {92}},\ \bibinfo
  {pages} {065003} (\bibinfo {year} {2015})},\ \Eprint
  {http://arxiv.org/abs/1504.07619} {arXiv:1504.07619 [hep-ph]} \BibitemShut
  {NoStop}%
\bibitem [{\citenamefont {Ozaki}\ \emph {et~al.}(2016)\citenamefont {Ozaki},
  \citenamefont {Itakura},\ and\ \citenamefont {Kuramoto}}]{Ozaki:2015sya}%
  \BibitemOpen
  \bibfield  {author} {\bibinfo {author} {\bibfnamefont {Sho}\ \bibnamefont
  {Ozaki}}, \bibinfo {author} {\bibfnamefont {Kazunori}\ \bibnamefont
  {Itakura}}, \ and\ \bibinfo {author} {\bibfnamefont {Yoshio}\ \bibnamefont
  {Kuramoto}},\ }\bibfield  {title} {\enquote {\bibinfo {title} {{Magnetically
  induced QCD Kondo effect}},}\ }\href
  {https://doi.org/10.1103/PhysRevD.94.074013} {\bibfield  {journal} {\bibinfo
  {journal} {Phys. Rev. D}\ }\textbf {\bibinfo {volume} {94}},\ \bibinfo
  {pages} {074013} (\bibinfo {year} {2016})},\ \Eprint
  {http://arxiv.org/abs/1509.06966} {arXiv:1509.06966 [hep-ph]} \BibitemShut
  {NoStop}%
\bibitem [{\citenamefont {Yasui}\ \emph {et~al.}(2019)\citenamefont {Yasui},
  \citenamefont {Suzuki},\ and\ \citenamefont {Itakura}}]{Yasui:2016svc}%
  \BibitemOpen
  \bibfield  {author} {\bibinfo {author} {\bibfnamefont {Shigehiro}\
  \bibnamefont {Yasui}}, \bibinfo {author} {\bibfnamefont {Kei}\ \bibnamefont
  {Suzuki}}, \ and\ \bibinfo {author} {\bibfnamefont {Kazunori}\ \bibnamefont
  {Itakura}},\ }\bibfield  {title} {\enquote {\bibinfo {title} {{Kondo phase
  diagram of quark matter}},}\ }\href
  {https://doi.org/10.1016/j.nuclphysa.2018.12.025} {\bibfield  {journal}
  {\bibinfo  {journal} {Nucl. Phys. A}\ }\textbf {\bibinfo {volume} {983}},\
  \bibinfo {pages} {90--102} (\bibinfo {year} {2019})},\ \Eprint
  {http://arxiv.org/abs/1604.07208} {arXiv:1604.07208 [hep-ph]} \BibitemShut
  {NoStop}%
\bibitem [{\citenamefont {Yasui}(2017)}]{Yasui:2016yet}%
  \BibitemOpen
  \bibfield  {author} {\bibinfo {author} {\bibfnamefont {Shigehiro}\
  \bibnamefont {Yasui}},\ }\bibfield  {title} {\enquote {\bibinfo {title}
  {{Kondo cloud of single heavy quark in cold and dense matter}},}\ }\href
  {https://doi.org/10.1016/j.physletb.2017.08.066} {\bibfield  {journal}
  {\bibinfo  {journal} {Phys. Lett. B}\ }\textbf {\bibinfo {volume} {773}},\
  \bibinfo {pages} {428--434} (\bibinfo {year} {2017})},\ \Eprint
  {http://arxiv.org/abs/1608.06450} {arXiv:1608.06450 [hep-ph]} \BibitemShut
  {NoStop}%
\bibitem [{\citenamefont {Kimura}\ and\ \citenamefont
  {Ozaki}(2017)}]{Kimura:2016zyv}%
  \BibitemOpen
  \bibfield  {author} {\bibinfo {author} {\bibfnamefont {Taro}\ \bibnamefont
  {Kimura}}\ and\ \bibinfo {author} {\bibfnamefont {Sho}\ \bibnamefont
  {Ozaki}},\ }\bibfield  {title} {\enquote {\bibinfo {title} {{Fermi/non-Fermi
  mixing in SU($N$) Kondo effect}},}\ }\href
  {https://doi.org/10.7566/JPSJ.86.084703} {\bibfield  {journal} {\bibinfo
  {journal} {J. Phys. Soc. Jap.}\ }\textbf {\bibinfo {volume} {86}},\ \bibinfo
  {pages} {084703} (\bibinfo {year} {2017})},\ \Eprint
  {http://arxiv.org/abs/1611.07284} {arXiv:1611.07284 [cond-mat.str-el]}
  \BibitemShut {NoStop}%
\bibitem [{\citenamefont {Yasui}\ \emph {et~al.}(2017)\citenamefont {Yasui},
  \citenamefont {Suzuki},\ and\ \citenamefont {Itakura}}]{Yasui:2017izi}%
  \BibitemOpen
  \bibfield  {author} {\bibinfo {author} {\bibfnamefont {Shigehiro}\
  \bibnamefont {Yasui}}, \bibinfo {author} {\bibfnamefont {Kei}\ \bibnamefont
  {Suzuki}}, \ and\ \bibinfo {author} {\bibfnamefont {Kazunori}\ \bibnamefont
  {Itakura}},\ }\bibfield  {title} {\enquote {\bibinfo {title} {{Topology and
  stability of the Kondo phase in quark matter}},}\ }\href
  {https://doi.org/10.1103/PhysRevD.96.014016} {\bibfield  {journal} {\bibinfo
  {journal} {Phys. Rev. D}\ }\textbf {\bibinfo {volume} {96}},\ \bibinfo
  {pages} {014016} (\bibinfo {year} {2017})},\ \Eprint
  {http://arxiv.org/abs/1703.04124} {arXiv:1703.04124 [hep-ph]} \BibitemShut
  {NoStop}%
\bibitem [{\citenamefont {Suzuki}\ \emph {et~al.}(2017)\citenamefont {Suzuki},
  \citenamefont {Yasui},\ and\ \citenamefont {Itakura}}]{Suzuki:2017gde}%
  \BibitemOpen
  \bibfield  {author} {\bibinfo {author} {\bibfnamefont {Kei}\ \bibnamefont
  {Suzuki}}, \bibinfo {author} {\bibfnamefont {Shigehiro}\ \bibnamefont
  {Yasui}}, \ and\ \bibinfo {author} {\bibfnamefont {Kazunori}\ \bibnamefont
  {Itakura}},\ }\bibfield  {title} {\enquote {\bibinfo {title} {{Interplay
  between chiral symmetry breaking and the QCD Kondo effect}},}\ }\href
  {https://doi.org/10.1103/PhysRevD.96.114007} {\bibfield  {journal} {\bibinfo
  {journal} {Phys. Rev. D}\ }\textbf {\bibinfo {volume} {96}},\ \bibinfo
  {pages} {114007} (\bibinfo {year} {2017})},\ \Eprint
  {http://arxiv.org/abs/1708.06930} {arXiv:1708.06930 [hep-ph]} \BibitemShut
  {NoStop}%
\bibitem [{\citenamefont {Yasui}\ and\ \citenamefont
  {Ozaki}(2017)}]{Yasui:2017bey}%
  \BibitemOpen
  \bibfield  {author} {\bibinfo {author} {\bibfnamefont {Shigehiro}\
  \bibnamefont {Yasui}}\ and\ \bibinfo {author} {\bibfnamefont {Sho}\
  \bibnamefont {Ozaki}},\ }\bibfield  {title} {\enquote {\bibinfo {title}
  {{Transport coefficients from the QCD Kondo effect}},}\ }\href
  {https://doi.org/10.1103/PhysRevD.96.114027} {\bibfield  {journal} {\bibinfo
  {journal} {Phys. Rev. D}\ }\textbf {\bibinfo {volume} {96}},\ \bibinfo
  {pages} {114027} (\bibinfo {year} {2017})},\ \Eprint
  {http://arxiv.org/abs/1710.03434} {arXiv:1710.03434 [hep-ph]} \BibitemShut
  {NoStop}%
\bibitem [{\citenamefont {Kimura}\ and\ \citenamefont
  {Ozaki}(2019)}]{Kimura:2018vxj}%
  \BibitemOpen
  \bibfield  {author} {\bibinfo {author} {\bibfnamefont {Taro}\ \bibnamefont
  {Kimura}}\ and\ \bibinfo {author} {\bibfnamefont {Sho}\ \bibnamefont
  {Ozaki}},\ }\bibfield  {title} {\enquote {\bibinfo {title} {{Conformal field
  theory analysis of the QCD Kondo effect}},}\ }\href
  {https://doi.org/10.1103/PhysRevD.99.014040} {\bibfield  {journal} {\bibinfo
  {journal} {Phys. Rev. D}\ }\textbf {\bibinfo {volume} {99}},\ \bibinfo
  {pages} {014040} (\bibinfo {year} {2019})},\ \Eprint
  {http://arxiv.org/abs/1806.06486} {arXiv:1806.06486 [hep-ph]} \BibitemShut
  {NoStop}%
\bibitem [{\citenamefont {Fariello}\ \emph {et~al.}(2019)\citenamefont
  {Fariello}, \citenamefont {Mac\'{i}as},\ and\ \citenamefont
  {Navarra}}]{Macias:2019vbl}%
  \BibitemOpen
  \bibfield  {author} {\bibinfo {author} {\bibfnamefont {R.}~\bibnamefont
  {Fariello}}, \bibinfo {author} {\bibfnamefont {Juan~C.}\ \bibnamefont
  {Mac\'{i}as}}, \ and\ \bibinfo {author} {\bibfnamefont {F.S.}\ \bibnamefont
  {Navarra}},\ }\bibfield  {title} {\enquote {\bibinfo {title} {{The QCD Kondo
  phase in quark stars}},}\ }\href@noop {} {\  (\bibinfo {year} {2019})},\
  \Eprint {http://arxiv.org/abs/1901.01623} {arXiv:1901.01623 [nucl-th]}
  \BibitemShut {NoStop}%
\bibitem [{\citenamefont {Hattori}\ \emph {et~al.}(2019)\citenamefont
  {Hattori}, \citenamefont {Huang},\ and\ \citenamefont
  {Pisarski}}]{Hattori:2019zig}%
  \BibitemOpen
  \bibfield  {author} {\bibinfo {author} {\bibfnamefont {Koichi}\ \bibnamefont
  {Hattori}}, \bibinfo {author} {\bibfnamefont {Xu-Guang}\ \bibnamefont
  {Huang}}, \ and\ \bibinfo {author} {\bibfnamefont {Robert~D.}\ \bibnamefont
  {Pisarski}},\ }\bibfield  {title} {\enquote {\bibinfo {title} {{Emergent QCD
  Kondo effect in two-flavor color superconducting phase}},}\ }\href
  {https://doi.org/10.1103/PhysRevD.99.094044} {\bibfield  {journal} {\bibinfo
  {journal} {Phys. Rev. D}\ }\textbf {\bibinfo {volume} {99}},\ \bibinfo
  {pages} {094044} (\bibinfo {year} {2019})},\ \Eprint
  {http://arxiv.org/abs/1903.10953} {arXiv:1903.10953 [hep-ph]} \BibitemShut
  {NoStop}%
\bibitem [{\citenamefont {Suenaga}\ \emph
  {et~al.}(2020{\natexlab{a}})\citenamefont {Suenaga}, \citenamefont {Suzuki},\
  and\ \citenamefont {Yasui}}]{Suenaga:2019car}%
  \BibitemOpen
  \bibfield  {author} {\bibinfo {author} {\bibfnamefont {Daiki}\ \bibnamefont
  {Suenaga}}, \bibinfo {author} {\bibfnamefont {Kei}\ \bibnamefont {Suzuki}}, \
  and\ \bibinfo {author} {\bibfnamefont {Shigehiro}\ \bibnamefont {Yasui}},\
  }\bibfield  {title} {\enquote {\bibinfo {title} {{QCD Kondo excitons}},}\
  }\href {https://doi.org/10.1103/PhysRevResearch.2.023066} {\bibfield
  {journal} {\bibinfo  {journal} {Phys. Rev. Research}\ }\textbf {\bibinfo
  {volume} {2}},\ \bibinfo {pages} {023066} (\bibinfo {year}
  {2020}{\natexlab{a}})},\ \Eprint {http://arxiv.org/abs/1909.07573}
  {arXiv:1909.07573 [nucl-th]} \BibitemShut {NoStop}%
\bibitem [{\citenamefont {Suenaga}\ \emph
  {et~al.}(2020{\natexlab{b}})\citenamefont {Suenaga}, \citenamefont {Suzuki},
  \citenamefont {Araki},\ and\ \citenamefont {Yasui}}]{Suenaga:2019jqu}%
  \BibitemOpen
  \bibfield  {author} {\bibinfo {author} {\bibfnamefont {Daiki}\ \bibnamefont
  {Suenaga}}, \bibinfo {author} {\bibfnamefont {Kei}\ \bibnamefont {Suzuki}},
  \bibinfo {author} {\bibfnamefont {Yasufumi}\ \bibnamefont {Araki}}, \ and\
  \bibinfo {author} {\bibfnamefont {Shigehiro}\ \bibnamefont {Yasui}},\
  }\bibfield  {title} {\enquote {\bibinfo {title} {{Kondo effect driven by
  chirality imbalance}},}\ }\href
  {https://doi.org/10.1103/PhysRevResearch.2.023312} {\bibfield  {journal}
  {\bibinfo  {journal} {Phys. Rev. Research}\ }\textbf {\bibinfo {volume}
  {2}},\ \bibinfo {pages} {023312} (\bibinfo {year} {2020}{\natexlab{b}})},\
  \Eprint {http://arxiv.org/abs/1912.12669} {arXiv:1912.12669 [hep-ph]}
  \BibitemShut {NoStop}%
\bibitem [{\citenamefont {Kanazawa}(2020)}]{Kanazawa:2020xje}%
  \BibitemOpen
  \bibfield  {author} {\bibinfo {author} {\bibfnamefont {Takuya}\ \bibnamefont
  {Kanazawa}},\ }\bibfield  {title} {\enquote {\bibinfo {title} {{Random matrix
  model for the QCD Kondo effect}},}\ }\href@noop {} {\  (\bibinfo {year}
  {2020})},\ \Eprint {http://arxiv.org/abs/2006.00200} {arXiv:2006.00200
  [hep-th]} \BibitemShut {NoStop}%
\bibitem [{\citenamefont {Eichten}\ and\ \citenamefont
  {Hill}(1990)}]{Eichten:1989zv}%
  \BibitemOpen
  \bibfield  {author} {\bibinfo {author} {\bibfnamefont {Estia}\ \bibnamefont
  {Eichten}}\ and\ \bibinfo {author} {\bibfnamefont {Brian~Russell}\
  \bibnamefont {Hill}},\ }\bibfield  {title} {\enquote {\bibinfo {title} {{An
  effective field theory for the calculation of matrix elements involving heavy
  quarks}},}\ }\href {https://doi.org/10.1016/0370-2693(90)92049-O} {\bibfield
  {journal} {\bibinfo  {journal} {Phys. Lett. B}\ }\textbf {\bibinfo {volume}
  {234}},\ \bibinfo {pages} {511--516} (\bibinfo {year} {1990})}\BibitemShut
  {NoStop}%
\bibitem [{\citenamefont {Georgi}(1990)}]{Georgi:1990um}%
  \BibitemOpen
  \bibfield  {author} {\bibinfo {author} {\bibfnamefont {Howard}\ \bibnamefont
  {Georgi}},\ }\bibfield  {title} {\enquote {\bibinfo {title} {{An effective
  field theory for heavy quarks at low energies}},}\ }\href
  {https://doi.org/10.1016/0370-2693(90)91128-X} {\bibfield  {journal}
  {\bibinfo  {journal} {Phys. Lett. B}\ }\textbf {\bibinfo {volume} {240}},\
  \bibinfo {pages} {447--450} (\bibinfo {year} {1990})}\BibitemShut {NoStop}%
\bibitem [{\citenamefont {Neubert}(1994)}]{Neubert:1993mb}%
  \BibitemOpen
  \bibfield  {author} {\bibinfo {author} {\bibfnamefont {Matthias}\
  \bibnamefont {Neubert}},\ }\bibfield  {title} {\enquote {\bibinfo {title}
  {{Heavy-quark symmetry}},}\ }\href
  {https://doi.org/10.1016/0370-1573(94)90091-4} {\bibfield  {journal}
  {\bibinfo  {journal} {Phys. Rept.}\ }\textbf {\bibinfo {volume} {245}},\
  \bibinfo {pages} {259--395} (\bibinfo {year} {1994})},\ \Eprint
  {http://arxiv.org/abs/hep-ph/9306320} {arXiv:hep-ph/9306320 [hep-ph]}
  \BibitemShut {NoStop}%
\bibitem [{\citenamefont {Manohar}\ and\ \citenamefont
  {Wise}(2000)}]{Manohar:2000dt}%
  \BibitemOpen
  \bibfield  {author} {\bibinfo {author} {\bibfnamefont {Aneesh~V.}\
  \bibnamefont {Manohar}}\ and\ \bibinfo {author} {\bibfnamefont {Mark~B.}\
  \bibnamefont {Wise}},\ }\href {https://doi.org/10.1017/CBO9780511529351}
  {\emph {\bibinfo {title} {Heavy Quark Physics}}}\ (\bibinfo  {publisher}
  {Cambridge University Press, Cambridge, UK},\ \bibinfo {year}
  {2000})\BibitemShut {NoStop}%
\bibitem [{\citenamefont {Mannel}\ \emph {et~al.}(1992)\citenamefont {Mannel},
  \citenamefont {Roberts},\ and\ \citenamefont {Ryzak}}]{Mannel:1991mc}%
  \BibitemOpen
  \bibfield  {author} {\bibinfo {author} {\bibfnamefont {Thomas}\ \bibnamefont
  {Mannel}}, \bibinfo {author} {\bibfnamefont {Winston}\ \bibnamefont
  {Roberts}}, \ and\ \bibinfo {author} {\bibfnamefont {Zbigniew}\ \bibnamefont
  {Ryzak}},\ }\bibfield  {title} {\enquote {\bibinfo {title} {{A derivation of
  the heavy quark effective Lagrangian from QCD}},}\ }\href
  {https://doi.org/10.1016/0550-3213(92)90204-O} {\bibfield  {journal}
  {\bibinfo  {journal} {Nucl. Phys. B}\ }\textbf {\bibinfo {volume} {368}},\
  \bibinfo {pages} {204--217} (\bibinfo {year} {1992})}\BibitemShut {NoStop}%
\bibitem [{\citenamefont {K{\"o}rner}\ and\ \citenamefont
  {Thompson}(1991)}]{Korner:1991kf}%
  \BibitemOpen
  \bibfield  {author} {\bibinfo {author} {\bibfnamefont {J.~G.}\ \bibnamefont
  {K{\"o}rner}}\ and\ \bibinfo {author} {\bibfnamefont {George}\ \bibnamefont
  {Thompson}},\ }\bibfield  {title} {\enquote {\bibinfo {title} {{The heavy
  mass limit in field theory and the heavy quark effective theory}},}\ }\href
  {https://doi.org/10.1016/0370-2693(91)90725-6} {\bibfield  {journal}
  {\bibinfo  {journal} {Phys. Lett. B}\ }\textbf {\bibinfo {volume} {264}},\
  \bibinfo {pages} {185--192} (\bibinfo {year} {1991})}\BibitemShut {NoStop}%
\bibitem [{\citenamefont {Balk}\ \emph {et~al.}(1994)\citenamefont {Balk},
  \citenamefont {K{\"o}rner},\ and\ \citenamefont {Pirjol}}]{Balk:1993ev}%
  \BibitemOpen
  \bibfield  {author} {\bibinfo {author} {\bibfnamefont {S.}~\bibnamefont
  {Balk}}, \bibinfo {author} {\bibfnamefont {J.~G.}\ \bibnamefont
  {K{\"o}rner}}, \ and\ \bibinfo {author} {\bibfnamefont {D.}~\bibnamefont
  {Pirjol}},\ }\bibfield  {title} {\enquote {\bibinfo {title} {{Quark effective
  theory at large orders in $1/m$}},}\ }\href
  {https://doi.org/10.1016/0550-3213(94)90211-9} {\bibfield  {journal}
  {\bibinfo  {journal} {Nucl. Phys. B}\ }\textbf {\bibinfo {volume} {428}},\
  \bibinfo {pages} {499--527} (\bibinfo {year} {1994})},\ \Eprint
  {http://arxiv.org/abs/hep-ph/9307230} {arXiv:hep-ph/9307230 [hep-ph]}
  \BibitemShut {NoStop}%
\bibitem [{\citenamefont {Das}\ and\ \citenamefont
  {Mathur}(1994)}]{Das:1993rf}%
  \BibitemOpen
  \bibfield  {author} {\bibinfo {author} {\bibfnamefont {Ashok}\ \bibnamefont
  {Das}}\ and\ \bibinfo {author} {\bibfnamefont {V.~S.}\ \bibnamefont
  {Mathur}},\ }\bibfield  {title} {\enquote {\bibinfo {title} {{More on
  symmetries in heavy quark effective theory}},}\ }\href
  {https://doi.org/10.1103/PhysRevD.49.2508} {\bibfield  {journal} {\bibinfo
  {journal} {Phys.\ Rev.\ D}\ }\textbf {\bibinfo {volume} {49}},\ \bibinfo
  {pages} {2508--2513} (\bibinfo {year} {1994})},\ \Eprint
  {http://arxiv.org/abs/hep-ph/9309306} {arXiv:hep-ph/9309306} \BibitemShut
  {NoStop}%
\bibitem [{\citenamefont {Das}(1994)}]{Das:1993jx}%
  \BibitemOpen
  \bibfield  {author} {\bibinfo {author} {\bibfnamefont {Ashok}\ \bibnamefont
  {Das}},\ }\bibfield  {title} {\enquote {\bibinfo {title} {{On the higher
  order corrections to heavy quark effective theory}},}\ }\href
  {https://doi.org/10.1142/S0217732394000368} {\bibfield  {journal} {\bibinfo
  {journal} {Mod. Phys. Lett. A}\ }\textbf {\bibinfo {volume} {9}},\ \bibinfo
  {pages} {341--354} (\bibinfo {year} {1994})},\ \Eprint
  {http://arxiv.org/abs/hep-ph/9310372} {arXiv:hep-ph/9310372 [hep-ph]}
  \BibitemShut {NoStop}%
\bibitem [{\citenamefont {Balk}\ \emph {et~al.}(1993)\citenamefont {Balk},
  \citenamefont {Ilakovac}, \citenamefont {K{\"o}rner},\ and\ \citenamefont
  {Pirjol}}]{Balk:1993cd}%
  \BibitemOpen
  \bibfield  {author} {\bibinfo {author} {\bibfnamefont {S.}~\bibnamefont
  {Balk}}, \bibinfo {author} {\bibfnamefont {A.}~\bibnamefont {Ilakovac}},
  \bibinfo {author} {\bibfnamefont {J.~G.}\ \bibnamefont {K{\"o}rner}}, \ and\
  \bibinfo {author} {\bibfnamefont {D.}~\bibnamefont {Pirjol}},\ }\bibfield
  {title} {\enquote {\bibinfo {title} {{Two different formulations of the heavy
  quark effective theory}},}\ }in\ \href@noop {} {\emph {\bibinfo {booktitle}
  {{Proceedings, 27th International Symposium Ahrenshoop on Theory of
  elementary particles: Wendisch-Rietz, Germany, September 7-11, 1993}}}}\
  (\bibinfo {year} {1993})\ pp.\ \bibinfo {pages} {315--326},\ \Eprint
  {http://arxiv.org/abs/hep-ph/9311348} {arXiv:hep-ph/9311348 [hep-ph]}
  \BibitemShut {NoStop}%
\bibitem [{\citenamefont {Blok}\ \emph {et~al.}(1997)\citenamefont {Blok},
  \citenamefont {K{\"o}rner}, \citenamefont {Pirjol},\ and\ \citenamefont
  {Rojas}}]{Blok:1996iz}%
  \BibitemOpen
  \bibfield  {author} {\bibinfo {author} {\bibfnamefont {B.}~\bibnamefont
  {Blok}}, \bibinfo {author} {\bibfnamefont {J.~G.}\ \bibnamefont
  {K{\"o}rner}}, \bibinfo {author} {\bibfnamefont {D.}~\bibnamefont {Pirjol}},
  \ and\ \bibinfo {author} {\bibfnamefont {J.~C.}\ \bibnamefont {Rojas}},\
  }\bibfield  {title} {\enquote {\bibinfo {title} {{Spectator effects in the
  heavy quark effective theory}},}\ }\href
  {https://doi.org/10.1016/S0550-3213(97)00202-2} {\bibfield  {journal}
  {\bibinfo  {journal} {Nucl. Phys. B}\ }\textbf {\bibinfo {volume} {496}},\
  \bibinfo {pages} {358--374} (\bibinfo {year} {1997})},\ \Eprint
  {http://arxiv.org/abs/hep-ph/9607233} {arXiv:hep-ph/9607233 [hep-ph]}
  \BibitemShut {NoStop}%
\bibitem [{\citenamefont {Holstein}(1997)}]{Holstein:1997}%
  \BibitemOpen
  \bibfield  {author} {\bibinfo {author} {\bibfnamefont {Barry~R.}\
  \bibnamefont {Holstein}},\ }\bibfield  {title} {\enquote {\bibinfo {title}
  {{Diagonalization of the Dirac equation: An alternative approach}},}\ }\href
  {https://doi.org/10.1119/1.18582} {\bibfield  {journal} {\bibinfo  {journal}
  {Am. J. Phys.}\ }\textbf {\bibinfo {volume} {65}},\ \bibinfo {pages} {519}
  (\bibinfo {year} {1997})}\BibitemShut {NoStop}%
\bibitem [{\citenamefont {G\aa{}rdestig}\ \emph {et~al.}(2007)\citenamefont
  {G\aa{}rdestig}, \citenamefont {Kubodera},\ and\ \citenamefont
  {Myhrer}}]{Gardestig:2007mk}%
  \BibitemOpen
  \bibfield  {author} {\bibinfo {author} {\bibfnamefont {A.}~\bibnamefont
  {G\aa{}rdestig}}, \bibinfo {author} {\bibfnamefont {K.}~\bibnamefont
  {Kubodera}}, \ and\ \bibinfo {author} {\bibfnamefont {F.}~\bibnamefont
  {Myhrer}},\ }\bibfield  {title} {\enquote {\bibinfo {title} {{Comparison of
  the heavy-fermion and Foldy-Wouthuysen formalisms at third order}},}\ }\href
  {https://doi.org/10.1103/PhysRevC.76.014005} {\bibfield  {journal} {\bibinfo
  {journal} {Phys. Rev. C}\ }\textbf {\bibinfo {volume} {76}},\ \bibinfo
  {pages} {014005} (\bibinfo {year} {2007})},\ \Eprint
  {http://arxiv.org/abs/0705.2885} {arXiv:0705.2885 [nucl-th]} \BibitemShut
  {NoStop}%
\bibitem [{\citenamefont {Foldy}\ and\ \citenamefont
  {Wouthuysen}(1950)}]{Foldy:1949wa}%
  \BibitemOpen
  \bibfield  {author} {\bibinfo {author} {\bibfnamefont {Leslie~L.}\
  \bibnamefont {Foldy}}\ and\ \bibinfo {author} {\bibfnamefont {Siegfried~A.}\
  \bibnamefont {Wouthuysen}},\ }\bibfield  {title} {\enquote {\bibinfo {title}
  {{On the Dirac Theory of Spin 1/2 Particles and Its Non-Relativistic
  Limit}},}\ }\href {https://doi.org/10.1103/PhysRev.78.29} {\bibfield
  {journal} {\bibinfo  {journal} {Phys. Rev.}\ }\textbf {\bibinfo {volume}
  {78}},\ \bibinfo {pages} {29--36} (\bibinfo {year} {1950})}\BibitemShut
  {NoStop}%
\bibitem [{\citenamefont {Tani}(1951)}]{Tani:1951}%
  \BibitemOpen
  \bibfield  {author} {\bibinfo {author} {\bibfnamefont {Smio}\ \bibnamefont
  {Tani}},\ }\bibfield  {title} {\enquote {\bibinfo {title} {{Connection
  between Particle Models and Field Theories, I: The Case Spin 1/2}},}\ }\href
  {https://doi.org/10.1143/ptp/6.3.267} {\bibfield  {journal} {\bibinfo
  {journal} {Prog. Theor. Phys.}\ }\textbf {\bibinfo {volume} {6}},\ \bibinfo
  {pages} {267--285} (\bibinfo {year} {1951})}\BibitemShut {NoStop}%
\bibitem [{\citenamefont {Feinberg}(1978)}]{Feinberg:1977rc}%
  \BibitemOpen
  \bibfield  {author} {\bibinfo {author} {\bibfnamefont {Frank~L.}\
  \bibnamefont {Feinberg}},\ }\bibfield  {title} {\enquote {\bibinfo {title}
  {{Hamiltonian formulation of non-Abelian gauge fields and nonrelativistic
  bound states}},}\ }\href {https://doi.org/10.1103/PhysRevD.17.2659}
  {\bibfield  {journal} {\bibinfo  {journal} {Phys. Rev. D}\ }\textbf {\bibinfo
  {volume} {17}},\ \bibinfo {pages} {2659} (\bibinfo {year}
  {1978})}\BibitemShut {NoStop}%
\bibitem [{\citenamefont {Dzero}\ \emph {et~al.}(2010)\citenamefont {Dzero},
  \citenamefont {Sun}, \citenamefont {Galitski},\ and\ \citenamefont
  {Coleman}}]{DzeroSunGalitskiColeman2010}%
  \BibitemOpen
  \bibfield  {author} {\bibinfo {author} {\bibfnamefont {Maxim}\ \bibnamefont
  {Dzero}}, \bibinfo {author} {\bibfnamefont {Kai}\ \bibnamefont {Sun}},
  \bibinfo {author} {\bibfnamefont {Victor}\ \bibnamefont {Galitski}}, \ and\
  \bibinfo {author} {\bibfnamefont {Piers}\ \bibnamefont {Coleman}},\
  }\bibfield  {title} {\enquote {\bibinfo {title} {{Topological Kondo
  Insulators}},}\ }\href {https://doi.org/10.1103/PhysRevLett.104.106408}
  {\bibfield  {journal} {\bibinfo  {journal} {Phys. Rev. Lett.}\ }\textbf
  {\bibinfo {volume} {104}},\ \bibinfo {pages} {106408} (\bibinfo {year}
  {2010})},\ \Eprint {http://arxiv.org/abs/0912.3750} {arXiv:0912.3750
  [cond-mat.str-el]} \BibitemShut {NoStop}%
\bibitem [{\citenamefont {Dzero}\ \emph {et~al.}(2012)\citenamefont {Dzero},
  \citenamefont {Sun}, \citenamefont {Coleman},\ and\ \citenamefont
  {Galitski}}]{DzeroSunColemanGalitski2012}%
  \BibitemOpen
  \bibfield  {author} {\bibinfo {author} {\bibfnamefont {Maxim}\ \bibnamefont
  {Dzero}}, \bibinfo {author} {\bibfnamefont {Kai}\ \bibnamefont {Sun}},
  \bibinfo {author} {\bibfnamefont {Piers}\ \bibnamefont {Coleman}}, \ and\
  \bibinfo {author} {\bibfnamefont {Victor}\ \bibnamefont {Galitski}},\
  }\bibfield  {title} {\enquote {\bibinfo {title} {{Theory of topological Kondo
  insulators}},}\ }\href {https://doi.org/10.1103/PhysRevB.85.045130}
  {\bibfield  {journal} {\bibinfo  {journal} {Phys. Rev. B}\ }\textbf {\bibinfo
  {volume} {85}},\ \bibinfo {pages} {045130} (\bibinfo {year} {2012})},\
  \Eprint {http://arxiv.org/abs/1108.3371} {arXiv:1108.3371 [cond-mat.str-el]}
  \BibitemShut {NoStop}%
\bibitem [{\citenamefont {Tran}\ \emph {et~al.}(2012)\citenamefont {Tran},
  \citenamefont {Takimoto},\ and\ \citenamefont {Kim}}]{TranTakimotoKim2012}%
  \BibitemOpen
  \bibfield  {author} {\bibinfo {author} {\bibfnamefont {Minh-Tien}\
  \bibnamefont {Tran}}, \bibinfo {author} {\bibfnamefont {Tetsuya}\
  \bibnamefont {Takimoto}}, \ and\ \bibinfo {author} {\bibfnamefont {Ki-Seok}\
  \bibnamefont {Kim}},\ }\bibfield  {title} {\enquote {\bibinfo {title} {{Phase
  diagram for a topological Kondo insulating system}},}\ }\href
  {https://doi.org/10.1103/PhysRevB.85.125128} {\bibfield  {journal} {\bibinfo
  {journal} {Phys. Rev. B}\ }\textbf {\bibinfo {volume} {85}},\ \bibinfo
  {pages} {125128} (\bibinfo {year} {2012})},\ \Eprint
  {http://arxiv.org/abs/1109.5788} {arXiv:1109.5788 [cond-mat.str-el]}
  \BibitemShut {NoStop}%
\bibitem [{\citenamefont {Takimoto}(2011)}]{Takimoto:2011}%
  \BibitemOpen
  \bibfield  {author} {\bibinfo {author} {\bibfnamefont {Tetsuya}\ \bibnamefont
  {Takimoto}},\ }\bibfield  {title} {\enquote {\bibinfo {title} {{SmB$_6$: A
  Promising Candidate for a Topological Insulator}},}\ }\href
  {https://doi.org/10.1143/JPSJ.80.123710} {\bibfield  {journal} {\bibinfo
  {journal} {J. Phys. Soc. Jpn.}\ }\textbf {\bibinfo {volume} {80}},\ \bibinfo
  {pages} {123710} (\bibinfo {year} {2011})}\BibitemShut {NoStop}%
\bibitem [{\citenamefont {Werner}\ and\ \citenamefont
  {Assaad}(2013)}]{WernerAssaad2013}%
  \BibitemOpen
  \bibfield  {author} {\bibinfo {author} {\bibfnamefont {Jan}\ \bibnamefont
  {Werner}}\ and\ \bibinfo {author} {\bibfnamefont {Fakher~F.}\ \bibnamefont
  {Assaad}},\ }\bibfield  {title} {\enquote {\bibinfo {title}
  {{Interaction-driven transition between topological states in a Kondo
  insulator}},}\ }\href {https://doi.org/10.1103/PhysRevB.88.035113} {\bibfield
   {journal} {\bibinfo  {journal} {Phys. Rev. B}\ }\textbf {\bibinfo {volume}
  {88}},\ \bibinfo {pages} {035113} (\bibinfo {year} {2013})},\ \Eprint
  {http://arxiv.org/abs/1302.1874} {arXiv:1302.1874 [cond-mat.str-el]}
  \BibitemShut {NoStop}%
\bibitem [{\citenamefont {Alexandrov}\ \emph {et~al.}(2013)\citenamefont
  {Alexandrov}, \citenamefont {Dzero},\ and\ \citenamefont
  {Coleman}}]{Alexandrov:2013}%
  \BibitemOpen
  \bibfield  {author} {\bibinfo {author} {\bibfnamefont {Victor}\ \bibnamefont
  {Alexandrov}}, \bibinfo {author} {\bibfnamefont {Maxim}\ \bibnamefont
  {Dzero}}, \ and\ \bibinfo {author} {\bibfnamefont {Piers}\ \bibnamefont
  {Coleman}},\ }\bibfield  {title} {\enquote {\bibinfo {title} {{Cubic
  Topological Kondo Insulators}},}\ }\href
  {https://doi.org/10.1103/PhysRevLett.111.226403} {\bibfield  {journal}
  {\bibinfo  {journal} {Phys. Rev. Lett.}\ }\textbf {\bibinfo {volume} {111}},\
  \bibinfo {pages} {226403} (\bibinfo {year} {2013})},\ \Eprint
  {http://arxiv.org/abs/1303.7224} {arXiv:1303.7224 [cond-mat.str-el]}
  \BibitemShut {NoStop}%
\bibitem [{\citenamefont {Werner}\ and\ \citenamefont
  {Assaad}(2014)}]{Werner:2014}%
  \BibitemOpen
  \bibfield  {author} {\bibinfo {author} {\bibfnamefont {Jan}\ \bibnamefont
  {Werner}}\ and\ \bibinfo {author} {\bibfnamefont {Fakher~F.}\ \bibnamefont
  {Assaad}},\ }\bibfield  {title} {\enquote {\bibinfo {title} {{Dynamically
  generated edge states in topological Kondo insulators}},}\ }\href
  {https://doi.org/10.1103/PhysRevB.89.245119} {\bibfield  {journal} {\bibinfo
  {journal} {Phys. Rev. B}\ }\textbf {\bibinfo {volume} {89}},\ \bibinfo
  {pages} {245119} (\bibinfo {year} {2014})},\ \Eprint
  {http://arxiv.org/abs/1311.3668} {arXiv:1311.3668 [cond-mat.str-el]}
  \BibitemShut {NoStop}%
\bibitem [{\citenamefont {Legner}\ \emph {et~al.}(2014)\citenamefont {Legner},
  \citenamefont {R\"uegg},\ and\ \citenamefont {Sigrist}}]{Legner:2014}%
  \BibitemOpen
  \bibfield  {author} {\bibinfo {author} {\bibfnamefont {Markus}\ \bibnamefont
  {Legner}}, \bibinfo {author} {\bibfnamefont {Andreas}\ \bibnamefont
  {R\"uegg}}, \ and\ \bibinfo {author} {\bibfnamefont {Manfred}\ \bibnamefont
  {Sigrist}},\ }\bibfield  {title} {\enquote {\bibinfo {title} {{Topological
  invariants, surface states, and interaction-driven phase transitions in
  correlated Kondo insulators with cubic symmetry}},}\ }\href
  {https://doi.org/10.1103/PhysRevB.89.085110} {\bibfield  {journal} {\bibinfo
  {journal} {Phys. Rev. B}\ }\textbf {\bibinfo {volume} {89}},\ \bibinfo
  {pages} {085110} (\bibinfo {year} {2014})},\ \Eprint
  {http://arxiv.org/abs/1312.3639} {arXiv:1312.3639 [cond-mat.str-el]}
  \BibitemShut {NoStop}%
\bibitem [{\citenamefont {Roy}\ \emph {et~al.}(2014)\citenamefont {Roy},
  \citenamefont {Sau}, \citenamefont {Dzero},\ and\ \citenamefont
  {Galitski}}]{Roy:2014}%
  \BibitemOpen
  \bibfield  {author} {\bibinfo {author} {\bibfnamefont {Bitan}\ \bibnamefont
  {Roy}}, \bibinfo {author} {\bibfnamefont {Jay~D.}\ \bibnamefont {Sau}},
  \bibinfo {author} {\bibfnamefont {Maxim}\ \bibnamefont {Dzero}}, \ and\
  \bibinfo {author} {\bibfnamefont {Victor}\ \bibnamefont {Galitski}},\
  }\bibfield  {title} {\enquote {\bibinfo {title} {{Surface theory of a family
  of topological Kondo insulators}},}\ }\href
  {https://doi.org/10.1103/PhysRevB.90.155314} {\bibfield  {journal} {\bibinfo
  {journal} {Phys. Rev. B}\ }\textbf {\bibinfo {volume} {90}},\ \bibinfo
  {pages} {155314} (\bibinfo {year} {2014})},\ \Eprint
  {http://arxiv.org/abs/1405.5526} {arXiv:1405.5526 [cond-mat.str-el]}
  \BibitemShut {NoStop}%
\bibitem [{\citenamefont {Alexandrov}\ \emph {et~al.}(2015)\citenamefont
  {Alexandrov}, \citenamefont {Coleman},\ and\ \citenamefont
  {Erten}}]{Alexandrov:2015}%
  \BibitemOpen
  \bibfield  {author} {\bibinfo {author} {\bibfnamefont {Victor}\ \bibnamefont
  {Alexandrov}}, \bibinfo {author} {\bibfnamefont {Piers}\ \bibnamefont
  {Coleman}}, \ and\ \bibinfo {author} {\bibfnamefont {Onur}\ \bibnamefont
  {Erten}},\ }\bibfield  {title} {\enquote {\bibinfo {title} {{Kondo Breakdown
  in Topological Kondo Insulators}},}\ }\href
  {https://doi.org/10.1103/PhysRevLett.114.177202} {\bibfield  {journal}
  {\bibinfo  {journal} {Phys. Rev. Lett.}\ }\textbf {\bibinfo {volume} {114}},\
  \bibinfo {pages} {177202} (\bibinfo {year} {2015})},\ \Eprint
  {http://arxiv.org/abs/1501.03031} {arXiv:1501.03031 [cond-mat.str-el]}
  \BibitemShut {NoStop}%
\bibitem [{\citenamefont {Legner}\ \emph {et~al.}(2015)\citenamefont {Legner},
  \citenamefont {R\"uegg},\ and\ \citenamefont {Sigrist}}]{Legner:2015}%
  \BibitemOpen
  \bibfield  {author} {\bibinfo {author} {\bibfnamefont {Markus}\ \bibnamefont
  {Legner}}, \bibinfo {author} {\bibfnamefont {Andreas}\ \bibnamefont
  {R\"uegg}}, \ and\ \bibinfo {author} {\bibfnamefont {Manfred}\ \bibnamefont
  {Sigrist}},\ }\bibfield  {title} {\enquote {\bibinfo {title} {{Surface-State
  Spin Textures and Mirror Chern Numbers in Topological Kondo Insulators}},}\
  }\href {https://doi.org/10.1103/PhysRevLett.115.156405} {\bibfield  {journal}
  {\bibinfo  {journal} {Phys. Rev. Lett.}\ }\textbf {\bibinfo {volume} {115}},\
  \bibinfo {pages} {156405} (\bibinfo {year} {2015})},\ \Eprint
  {http://arxiv.org/abs/1505.02987} {arXiv:1505.02987 [cond-mat.str-el]}
  \BibitemShut {NoStop}%
\bibitem [{\citenamefont {Dzero}\ \emph {et~al.}(2016)\citenamefont {Dzero},
  \citenamefont {Xia}, \citenamefont {Galitski},\ and\ \citenamefont
  {Coleman}}]{DzeroXiaGalitskiColeman2016}%
  \BibitemOpen
  \bibfield  {author} {\bibinfo {author} {\bibfnamefont {Maxim}\ \bibnamefont
  {Dzero}}, \bibinfo {author} {\bibfnamefont {Jing}\ \bibnamefont {Xia}},
  \bibinfo {author} {\bibfnamefont {Victor}\ \bibnamefont {Galitski}}, \ and\
  \bibinfo {author} {\bibfnamefont {Piers}\ \bibnamefont {Coleman}},\
  }\bibfield  {title} {\enquote {\bibinfo {title} {{Topological Kondo
  Insulators}},}\ }\href
  {https://doi.org/10.1146/annurev-conmatphys-031214-014749} {\bibfield
  {journal} {\bibinfo  {journal} {Annu. Rev. Condens. Matter Phys.}\ }\textbf
  {\bibinfo {volume} {7}},\ \bibinfo {pages} {249--280} (\bibinfo {year}
  {2016})},\ \Eprint {http://arxiv.org/abs/1506.05635} {arXiv:1506.05635
  [cond-mat.str-el]} \BibitemShut {NoStop}%
\bibitem [{\citenamefont {Baruselli}\ and\ \citenamefont
  {Vojta}(2016)}]{Baruselli:2016}%
  \BibitemOpen
  \bibfield  {author} {\bibinfo {author} {\bibfnamefont {Pier~Paolo}\
  \bibnamefont {Baruselli}}\ and\ \bibinfo {author} {\bibfnamefont {Matthias}\
  \bibnamefont {Vojta}},\ }\bibfield  {title} {\enquote {\bibinfo {title}
  {{Cotunneling into a Kondo lattice with odd hybridization}},}\ }\href
  {https://doi.org/10.1103/PhysRevB.93.235111} {\bibfield  {journal} {\bibinfo
  {journal} {Phys. Rev. B}\ }\textbf {\bibinfo {volume} {93}},\ \bibinfo
  {pages} {235111} (\bibinfo {year} {2016})},\ \Eprint
  {http://arxiv.org/abs/1603.00620} {arXiv:1603.00620 [cond-mat.str-el]}
  \BibitemShut {NoStop}%
\bibitem [{\citenamefont {Takasan}\ \emph {et~al.}(2017)\citenamefont
  {Takasan}, \citenamefont {Nakagawa},\ and\ \citenamefont
  {Kawakami}}]{Takasan:2017}%
  \BibitemOpen
  \bibfield  {author} {\bibinfo {author} {\bibfnamefont {Kazuaki}\ \bibnamefont
  {Takasan}}, \bibinfo {author} {\bibfnamefont {Masaya}\ \bibnamefont
  {Nakagawa}}, \ and\ \bibinfo {author} {\bibfnamefont {Norio}\ \bibnamefont
  {Kawakami}},\ }\bibfield  {title} {\enquote {\bibinfo {title}
  {{Laser-irradiated Kondo insulators: Controlling the Kondo effect and
  topological phases}},}\ }\href {https://doi.org/10.1103/PhysRevB.96.115120}
  {\bibfield  {journal} {\bibinfo  {journal} {Phys. Rev. B}\ }\textbf {\bibinfo
  {volume} {96}},\ \bibinfo {pages} {115120} (\bibinfo {year} {2017})},\
  \Eprint {http://arxiv.org/abs/1706.06114} {arXiv:1706.06114
  [cond-mat.str-el]} \BibitemShut {NoStop}%
\bibitem [{\citenamefont {Chang}\ and\ \citenamefont
  {Coleman}(2018)}]{Chang:2017}%
  \BibitemOpen
  \bibfield  {author} {\bibinfo {author} {\bibfnamefont {Po-Yao}\ \bibnamefont
  {Chang}}\ and\ \bibinfo {author} {\bibfnamefont {Piers}\ \bibnamefont
  {Coleman}},\ }\bibfield  {title} {\enquote {\bibinfo {title}
  {{Parity-violating hybridization in heavy Weyl semimetals}},}\ }\href
  {https://doi.org/10.1103/PhysRevB.97.155134} {\bibfield  {journal} {\bibinfo
  {journal} {Phys. Rev. B}\ }\textbf {\bibinfo {volume} {97}},\ \bibinfo
  {pages} {155134} (\bibinfo {year} {2018})},\ \Eprint
  {http://arxiv.org/abs/1710.09928} {arXiv:1710.09928 [cond-mat.str-el]}
  \BibitemShut {NoStop}%
\bibitem [{\citenamefont {Peters}\ \emph {et~al.}(2018)\citenamefont {Peters},
  \citenamefont {Yoshida},\ and\ \citenamefont {Kawakami}}]{Peters:2018}%
  \BibitemOpen
  \bibfield  {author} {\bibinfo {author} {\bibfnamefont {Robert}\ \bibnamefont
  {Peters}}, \bibinfo {author} {\bibfnamefont {Tsuneya}\ \bibnamefont
  {Yoshida}}, \ and\ \bibinfo {author} {\bibfnamefont {Norio}\ \bibnamefont
  {Kawakami}},\ }\bibfield  {title} {\enquote {\bibinfo {title} {{Magnetic
  states in a three-dimensional topological Kondo insulator}},}\ }\href
  {https://doi.org/10.1103/PhysRevB.98.075104} {\bibfield  {journal} {\bibinfo
  {journal} {Phys. Rev. B}\ }\textbf {\bibinfo {volume} {98}},\ \bibinfo
  {pages} {075104} (\bibinfo {year} {2018})},\ \Eprint
  {http://arxiv.org/abs/1804.04802} {arXiv:1804.04802 [cond-mat.str-el]}
  \BibitemShut {NoStop}%
\bibitem [{\citenamefont {Li}\ \emph {et~al.}(2018{\natexlab{b}})\citenamefont
  {Li}, \citenamefont {Zhong}, \citenamefont {Liu}, \citenamefont {Luo},\ and\
  \citenamefont {Song}}]{Li_1806.05578}%
  \BibitemOpen
  \bibfield  {author} {\bibinfo {author} {\bibfnamefont {Huan}\ \bibnamefont
  {Li}}, \bibinfo {author} {\bibfnamefont {Yin}\ \bibnamefont {Zhong}},
  \bibinfo {author} {\bibfnamefont {Yu}~\bibnamefont {Liu}}, \bibinfo {author}
  {\bibfnamefont {Hong-Gang}\ \bibnamefont {Luo}}, \ and\ \bibinfo {author}
  {\bibfnamefont {Hai-Feng}\ \bibnamefont {Song}},\ }\bibfield  {title}
  {\enquote {\bibinfo {title} {{$\mathcal {Z} _2$ classification for a novel
  antiferromagnetic topological insulating phase in three-dimensional
  topological Kondo insulator}},}\ }\href
  {https://doi.org/10.1088/1361-648x/aae17b} {\bibfield  {journal} {\bibinfo
  {journal} {J. Phys.: Condens. Matter}\ }\textbf {\bibinfo {volume} {30}},\
  \bibinfo {pages} {435601} (\bibinfo {year} {2018}{\natexlab{b}})},\ \Eprint
  {http://arxiv.org/abs/1806.05578} {arXiv:1806.05578 [cond-mat.str-el]}
  \BibitemShut {NoStop}%
\bibitem [{\citenamefont {Li}\ \emph {et~al.}(2018{\natexlab{c}})\citenamefont
  {Li}, \citenamefont {Wang}, \citenamefont {Zheng}, \citenamefont {Liu},\ and\
  \citenamefont {Zhong}}]{Li_1809.09867}%
  \BibitemOpen
  \bibfield  {author} {\bibinfo {author} {\bibfnamefont {Huan}\ \bibnamefont
  {Li}}, \bibinfo {author} {\bibfnamefont {Zhi-Yong}\ \bibnamefont {Wang}},
  \bibinfo {author} {\bibfnamefont {Xiao-Jun}\ \bibnamefont {Zheng}}, \bibinfo
  {author} {\bibfnamefont {Yu}~\bibnamefont {Liu}}, \ and\ \bibinfo {author}
  {\bibfnamefont {Yin}\ \bibnamefont {Zhong}},\ }\bibfield  {title} {\enquote
  {\bibinfo {title} {{Magnetic and topological transitions in three-dimensional
  topological Kondo insulator}},}\ }\href
  {https://doi.org/10.1088/0256-307x/35/12/127501} {\bibfield  {journal}
  {\bibinfo  {journal} {Chin. Phys. Lett.}\ }\textbf {\bibinfo {volume} {35}},\
  \bibinfo {pages} {127501} (\bibinfo {year} {2018}{\natexlab{c}})},\ \Eprint
  {http://arxiv.org/abs/1809.09867} {arXiv:1809.09867 [cond-mat.str-el]}
  \BibitemShut {NoStop}%
\bibitem [{\citenamefont {Lu}\ \emph {et~al.}(2019)\citenamefont {Lu},
  \citenamefont {Chou}, \citenamefont {Chung},\ and\ \citenamefont
  {Mou}}]{Lu:2019}%
  \BibitemOpen
  \bibfield  {author} {\bibinfo {author} {\bibfnamefont {Yen-Wen}\ \bibnamefont
  {Lu}}, \bibinfo {author} {\bibfnamefont {Po-Hao}\ \bibnamefont {Chou}},
  \bibinfo {author} {\bibfnamefont {Chung-Hou}\ \bibnamefont {Chung}}, \ and\
  \bibinfo {author} {\bibfnamefont {Chung-Yu}\ \bibnamefont {Mou}},\ }\bibfield
   {title} {\enquote {\bibinfo {title} {{Tunable topological semimetallic
  phases in Kondo lattice systems}},}\ }\href
  {https://doi.org/10.1103/PhysRevB.99.035141} {\bibfield  {journal} {\bibinfo
  {journal} {Phys. Rev. B}\ }\textbf {\bibinfo {volume} {99}},\ \bibinfo
  {pages} {035141} (\bibinfo {year} {2019})},\ \Eprint
  {http://arxiv.org/abs/1901.03191} {arXiv:1901.03191 [cond-mat.str-el]}
  \BibitemShut {NoStop}%
\bibitem [{\citenamefont {Peters}\ \emph {et~al.}(2019)\citenamefont {Peters},
  \citenamefont {Yoshida},\ and\ \citenamefont {Kawakami}}]{Peters:2019}%
  \BibitemOpen
  \bibfield  {author} {\bibinfo {author} {\bibfnamefont {Robert}\ \bibnamefont
  {Peters}}, \bibinfo {author} {\bibfnamefont {Tsuneya}\ \bibnamefont
  {Yoshida}}, \ and\ \bibinfo {author} {\bibfnamefont {Norio}\ \bibnamefont
  {Kawakami}},\ }\bibfield  {title} {\enquote {\bibinfo {title} {{Quantum
  oscillations in strongly correlated topological Kondo insulators}},}\ }\href
  {https://doi.org/10.1103/PhysRevB.100.085124} {\bibfield  {journal} {\bibinfo
   {journal} {Phys. Rev. B}\ }\textbf {\bibinfo {volume} {100}},\ \bibinfo
  {pages} {085124} (\bibinfo {year} {2019})},\ \Eprint
  {http://arxiv.org/abs/1901.05099} {arXiv:1901.05099 [cond-mat.str-el]}
  \BibitemShut {NoStop}%
\bibitem [{\citenamefont {Zhuang}\ \emph {et~al.}(2019)\citenamefont {Zhuang},
  \citenamefont {Zheng}, \citenamefont {Wang}, \citenamefont {Ming},
  \citenamefont {Li}, \citenamefont {Liu},\ and\ \citenamefont
  {Song}}]{Zhuang:2019}%
  \BibitemOpen
  \bibfield  {author} {\bibinfo {author} {\bibfnamefont {Jia-Tao}\ \bibnamefont
  {Zhuang}}, \bibinfo {author} {\bibfnamefont {Xiao-Jun}\ \bibnamefont
  {Zheng}}, \bibinfo {author} {\bibfnamefont {Zhi-Yong}\ \bibnamefont {Wang}},
  \bibinfo {author} {\bibfnamefont {Xing}\ \bibnamefont {Ming}}, \bibinfo
  {author} {\bibfnamefont {Huan}\ \bibnamefont {Li}}, \bibinfo {author}
  {\bibfnamefont {Yu}~\bibnamefont {Liu}}, \ and\ \bibinfo {author}
  {\bibfnamefont {Hai-Feng}\ \bibnamefont {Song}},\ }\bibfield  {title}
  {\enquote {\bibinfo {title} {{Valence transition in topological Kondo
  insulator}},}\ }\href {https://doi.org/10.1088/1361-648X/ab4625} {\bibfield
  {journal} {\bibinfo  {journal} {J. Phys. Condens. Matter}\ }\textbf {\bibinfo
  {volume} {32}},\ \bibinfo {pages} {035602} (\bibinfo {year} {2019})},\
  \Eprint {http://arxiv.org/abs/1908.00913} {arXiv:1908.00913
  [cond-mat.str-el]} \BibitemShut {NoStop}%
\bibitem [{\citenamefont {Tada}(2020)}]{Tada:2020}%
  \BibitemOpen
  \bibfield  {author} {\bibinfo {author} {\bibfnamefont {Yasuhiro}\
  \bibnamefont {Tada}},\ }\bibfield  {title} {\enquote {\bibinfo {title}
  {{Cyclotron resonance in Kondo insulator}},}\ }\href
  {https://doi.org/10.1103/PhysRevResearch.2.023194} {\bibfield  {journal}
  {\bibinfo  {journal} {Phys. Rev. Research}\ }\textbf {\bibinfo {volume}
  {2}},\ \bibinfo {pages} {023194} (\bibinfo {year} {2020})},\ \Eprint
  {http://arxiv.org/abs/2001.02819} {arXiv:2001.02819 [cond-mat.str-el]}
  \BibitemShut {NoStop}%
\bibitem [{\citenamefont {Alexandrov}\ and\ \citenamefont
  {Coleman}(2014)}]{Alexandrov:2014}%
  \BibitemOpen
  \bibfield  {author} {\bibinfo {author} {\bibfnamefont {Victor}\ \bibnamefont
  {Alexandrov}}\ and\ \bibinfo {author} {\bibfnamefont {Piers}\ \bibnamefont
  {Coleman}},\ }\bibfield  {title} {\enquote {\bibinfo {title} {{End states in
  a one-dimensional topological Kondo insulator in large-$N$ limit}},}\ }\href
  {https://doi.org/10.1103/PhysRevB.90.115147} {\bibfield  {journal} {\bibinfo
  {journal} {Phys. Rev. B}\ }\textbf {\bibinfo {volume} {90}},\ \bibinfo
  {pages} {115147} (\bibinfo {year} {2014})},\ \Eprint
  {http://arxiv.org/abs/1403.6819} {arXiv:1403.6819 [cond-mat.str-el]}
  \BibitemShut {NoStop}%
\bibitem [{\citenamefont {Lobos}\ \emph {et~al.}(2015)\citenamefont {Lobos},
  \citenamefont {Dobry},\ and\ \citenamefont {Galitski}}]{Lobos:2014}%
  \BibitemOpen
  \bibfield  {author} {\bibinfo {author} {\bibfnamefont {Alejandro~M.}\
  \bibnamefont {Lobos}}, \bibinfo {author} {\bibfnamefont {Ariel~O.}\
  \bibnamefont {Dobry}}, \ and\ \bibinfo {author} {\bibfnamefont {Victor}\
  \bibnamefont {Galitski}},\ }\bibfield  {title} {\enquote {\bibinfo {title}
  {{Magnetic End States in a Strongly Interacting One-Dimensional Topological
  Kondo Insulator}},}\ }\href {https://doi.org/10.1103/PhysRevX.5.021017}
  {\bibfield  {journal} {\bibinfo  {journal} {Phys. Rev. X}\ }\textbf {\bibinfo
  {volume} {5}},\ \bibinfo {pages} {021017} (\bibinfo {year} {2015})},\ \Eprint
  {http://arxiv.org/abs/1411.5357} {arXiv:1411.5357 [cond-mat.str-el]}
  \BibitemShut {NoStop}%
\bibitem [{\citenamefont {Mezio}\ \emph {et~al.}(2015)\citenamefont {Mezio},
  \citenamefont {Lobos}, \citenamefont {Dobry},\ and\ \citenamefont
  {Gazza}}]{Mezio:2015}%
  \BibitemOpen
  \bibfield  {author} {\bibinfo {author} {\bibfnamefont {Alejandro}\
  \bibnamefont {Mezio}}, \bibinfo {author} {\bibfnamefont {Alejandro~M.}\
  \bibnamefont {Lobos}}, \bibinfo {author} {\bibfnamefont {Ariel~O.}\
  \bibnamefont {Dobry}}, \ and\ \bibinfo {author} {\bibfnamefont {Claudio~J.}\
  \bibnamefont {Gazza}},\ }\bibfield  {title} {\enquote {\bibinfo {title}
  {{Haldane phase in one-dimensional topological Kondo insulators}},}\ }\href
  {https://doi.org/10.1103/PhysRevB.92.205128} {\bibfield  {journal} {\bibinfo
  {journal} {Phys. Rev. B}\ }\textbf {\bibinfo {volume} {92}},\ \bibinfo
  {pages} {205128} (\bibinfo {year} {2015})},\ \Eprint
  {http://arxiv.org/abs/1508.05927} {arXiv:1508.05927 [cond-mat.str-el]}
  \BibitemShut {NoStop}%
\bibitem [{\citenamefont {Hagym\'asi}\ and\ \citenamefont
  {Legeza}(2016)}]{Hagymasi:2016}%
  \BibitemOpen
  \bibfield  {author} {\bibinfo {author} {\bibfnamefont {I.}~\bibnamefont
  {Hagym\'asi}}\ and\ \bibinfo {author} {\bibfnamefont {\"O.}\ \bibnamefont
  {Legeza}},\ }\bibfield  {title} {\enquote {\bibinfo {title}
  {{Characterization of a correlated topological Kondo insulator in one
  dimension}},}\ }\href {https://doi.org/10.1103/PhysRevB.93.165104} {\bibfield
   {journal} {\bibinfo  {journal} {Phys. Rev. B}\ }\textbf {\bibinfo {volume}
  {93}},\ \bibinfo {pages} {165104} (\bibinfo {year} {2016})},\ \Eprint
  {http://arxiv.org/abs/1601.04606} {arXiv:1601.04606 [cond-mat.str-el]}
  \BibitemShut {NoStop}%
\bibitem [{\citenamefont {Zhong}\ \emph {et~al.}(2017)\citenamefont {Zhong},
  \citenamefont {Liu},\ and\ \citenamefont {Luo}}]{Zhong:2016}%
  \BibitemOpen
  \bibfield  {author} {\bibinfo {author} {\bibfnamefont {Yin}\ \bibnamefont
  {Zhong}}, \bibinfo {author} {\bibfnamefont {Yu}~\bibnamefont {Liu}}, \ and\
  \bibinfo {author} {\bibfnamefont {Hong-Gang}\ \bibnamefont {Luo}},\
  }\bibfield  {title} {\enquote {\bibinfo {title} {{Topological phase in 1D
  topological Kondo insulator: Z2 topological insulator, Haldane-like phase and
  Kondo breakdown}},}\ }\href {https://doi.org/10.1140/epjb/e2017-80102-0}
  {\bibfield  {journal} {\bibinfo  {journal} {Eur. Phys. J. B}\ }\textbf
  {\bibinfo {volume} {90}},\ \bibinfo {pages} {147} (\bibinfo {year} {2017})},\
  \Eprint {http://arxiv.org/abs/1612.09376} {arXiv:1612.09376
  [cond-mat.str-el]} \BibitemShut {NoStop}%
\bibitem [{\citenamefont {Lisandrini}\ \emph {et~al.}(2017)\citenamefont
  {Lisandrini}, \citenamefont {Lobos}, \citenamefont {Dobry},\ and\
  \citenamefont {Gazza}}]{Lisandrini:2017}%
  \BibitemOpen
  \bibfield  {author} {\bibinfo {author} {\bibfnamefont {Franco~T.}\
  \bibnamefont {Lisandrini}}, \bibinfo {author} {\bibfnamefont {Alejandro~M.}\
  \bibnamefont {Lobos}}, \bibinfo {author} {\bibfnamefont {Ariel~O.}\
  \bibnamefont {Dobry}}, \ and\ \bibinfo {author} {\bibfnamefont {Claudio~J.}\
  \bibnamefont {Gazza}},\ }\bibfield  {title} {\enquote {\bibinfo {title}
  {{Topological Kondo insulators in one dimension: Continuous Haldane-type
  ground-state evolution from the strongly interacting to the noninteracting
  limit}},}\ }\href {https://doi.org/10.1103/PhysRevB.96.075124} {\bibfield
  {journal} {\bibinfo  {journal} {Phys. Rev. B}\ }\textbf {\bibinfo {volume}
  {96}},\ \bibinfo {pages} {075124} (\bibinfo {year} {2017})},\ \Eprint
  {http://arxiv.org/abs/1704.02355} {arXiv:1704.02355 [cond-mat.str-el]}
  \BibitemShut {NoStop}%
\bibitem [{\citenamefont {Zhong}\ \emph {et~al.}(2019)\citenamefont {Zhong},
  \citenamefont {Wang}, \citenamefont {Liu}, \citenamefont {Song},
  \citenamefont {Liu},\ and\ \citenamefont {Luo}}]{Zhong:2017}%
  \BibitemOpen
  \bibfield  {author} {\bibinfo {author} {\bibfnamefont {Yin}\ \bibnamefont
  {Zhong}}, \bibinfo {author} {\bibfnamefont {Qin}\ \bibnamefont {Wang}},
  \bibinfo {author} {\bibfnamefont {Yu}~\bibnamefont {Liu}}, \bibinfo {author}
  {\bibfnamefont {Hai-Feng}\ \bibnamefont {Song}}, \bibinfo {author}
  {\bibfnamefont {Ke}~\bibnamefont {Liu}}, \ and\ \bibinfo {author}
  {\bibfnamefont {Hong-Gang}\ \bibnamefont {Luo}},\ }\bibfield  {title}
  {\enquote {\bibinfo {title} {{Finite temperature physics of 1D topological
  Kondo insulator: Stable Haldane phase, emergent energy scale and beyond}},}\
  }\href {https://doi.org/10.1007/s11467-018-0868-x} {\bibfield  {journal}
  {\bibinfo  {journal} {Front. Phys.}\ }\textbf {\bibinfo {volume} {14}},\
  \bibinfo {pages} {23602} (\bibinfo {year} {2019})},\ \Eprint
  {http://arxiv.org/abs/1710.02978} {arXiv:1710.02978 [cond-mat.str-el]}
  \BibitemShut {NoStop}%
\bibitem [{\citenamefont {Pillay}\ and\ \citenamefont
  {McCulloch}(2018)}]{Pillay:2018}%
  \BibitemOpen
  \bibfield  {author} {\bibinfo {author} {\bibfnamefont {Jason~C.}\
  \bibnamefont {Pillay}}\ and\ \bibinfo {author} {\bibfnamefont {Ian~P.}\
  \bibnamefont {McCulloch}},\ }\bibfield  {title} {\enquote {\bibinfo {title}
  {{Topological phase transition and the effect of Hubbard interactions on the
  one-dimensional topological Kondo insulator}},}\ }\href
  {https://doi.org/10.1103/PhysRevB.97.205133} {\bibfield  {journal} {\bibinfo
  {journal} {Phys. Rev. B}\ }\textbf {\bibinfo {volume} {97}},\ \bibinfo
  {pages} {205133} (\bibinfo {year} {2018})},\ \Eprint
  {http://arxiv.org/abs/1801.01279} {arXiv:1801.01279 [cond-mat.str-el]}
  \BibitemShut {NoStop}%
\bibitem [{\citenamefont {Hagym\'asi}\ \emph {et~al.}(2019)\citenamefont
  {Hagym\'asi}, \citenamefont {Hubig},\ and\ \citenamefont
  {Schollw\"ock}}]{Hagymasi:2018}%
  \BibitemOpen
  \bibfield  {author} {\bibinfo {author} {\bibfnamefont {I.}~\bibnamefont
  {Hagym\'asi}}, \bibinfo {author} {\bibfnamefont {C.}~\bibnamefont {Hubig}}, \
  and\ \bibinfo {author} {\bibfnamefont {U.}~\bibnamefont {Schollw\"ock}},\
  }\bibfield  {title} {\enquote {\bibinfo {title} {{Interaction quench and
  thermalization in a one-dimensional topological Kondo insulator}},}\ }\href
  {https://doi.org/10.1103/PhysRevB.99.075145} {\bibfield  {journal} {\bibinfo
  {journal} {Phys. Rev. B}\ }\textbf {\bibinfo {volume} {99}},\ \bibinfo
  {pages} {075145} (\bibinfo {year} {2019})},\ \Eprint
  {http://arxiv.org/abs/1810.09799} {arXiv:1810.09799 [cond-mat.str-el]}
  \BibitemShut {NoStop}%
\bibitem [{\citenamefont {Zhong}(2019)}]{Zhong:2019}%
  \BibitemOpen
  \bibfield  {author} {\bibinfo {author} {\bibfnamefont {Y}~\bibnamefont
  {Zhong}},\ }\bibfield  {title} {\enquote {\bibinfo {title} {{Understanding
  one-dimensional topological Kondo insulator: poor man's non-uniform
  antiferromagnetic mean-field theory versus quantum Monte Carlo
  simulation}},}\ }\href {https://doi.org/10.1140/epjb/e2019-100206-5}
  {\bibfield  {journal} {\bibinfo  {journal} {Eur. Phys. J. B}\ }\textbf
  {\bibinfo {volume} {92}},\ \bibinfo {pages} {178} (\bibinfo {year} {2019})},\
  \Eprint {http://arxiv.org/abs/1903.05294} {arXiv:1903.05294
  [cond-mat.str-el]} \BibitemShut {NoStop}%
\bibitem [{\citenamefont {Nambu}\ and\ \citenamefont
  {Jona-Lasinio}(1961{\natexlab{a}})}]{Nambu:1961tp}%
  \BibitemOpen
  \bibfield  {author} {\bibinfo {author} {\bibfnamefont {Y.}~\bibnamefont
  {Nambu}}\ and\ \bibinfo {author} {\bibfnamefont {G.}~\bibnamefont
  {Jona-Lasinio}},\ }\bibfield  {title} {\enquote {\bibinfo {title} {{Dynamical
  Model of Elementary Particles Based on an Analogy with Superconductivity.
  I}},}\ }\href {https://doi.org/10.1103/PhysRev.122.345} {\bibfield  {journal}
  {\bibinfo  {journal} {Phys. Rev.}\ }\textbf {\bibinfo {volume} {122}},\
  \bibinfo {pages} {345--358} (\bibinfo {year}
  {1961}{\natexlab{a}})}\BibitemShut {NoStop}%
\bibitem [{\citenamefont {Nambu}\ and\ \citenamefont
  {Jona-Lasinio}(1961{\natexlab{b}})}]{Nambu:1961fr}%
  \BibitemOpen
  \bibfield  {author} {\bibinfo {author} {\bibfnamefont {Y.}~\bibnamefont
  {Nambu}}\ and\ \bibinfo {author} {\bibfnamefont {G.}~\bibnamefont
  {Jona-Lasinio}},\ }\bibfield  {title} {\enquote {\bibinfo {title} {{Dynamical
  Model of Elementary Particles Based on an Analogy with Superconductivity.
  II}},}\ }\href {https://doi.org/10.1103/PhysRev.124.246} {\bibfield
  {journal} {\bibinfo  {journal} {Phys. Rev.}\ }\textbf {\bibinfo {volume}
  {124}},\ \bibinfo {pages} {246--254} (\bibinfo {year}
  {1961}{\natexlab{b}})}\BibitemShut {NoStop}%
\bibitem [{\citenamefont {Read}\ and\ \citenamefont {Newns}(1983)}]{Read:1983}%
  \BibitemOpen
  \bibfield  {author} {\bibinfo {author} {\bibfnamefont {N.}~\bibnamefont
  {Read}}\ and\ \bibinfo {author} {\bibfnamefont {D.~M.}\ \bibnamefont
  {Newns}},\ }\bibfield  {title} {\enquote {\bibinfo {title} {{On the solution
  of the Coqblin-Schreiffer Hamiltonian by the large-N expansion technique}},}\
  }\href {https://doi.org/10.1088/0022-3719/16/17/014} {\bibfield  {journal}
  {\bibinfo  {journal} {J. Phys. C}\ }\textbf {\bibinfo {volume} {16}},\
  \bibinfo {pages} {3273} (\bibinfo {year} {1983})}\BibitemShut {NoStop}%
\bibitem [{\citenamefont {Coleman}(1983)}]{Coleman:1983}%
  \BibitemOpen
  \bibfield  {author} {\bibinfo {author} {\bibfnamefont {Piers}\ \bibnamefont
  {Coleman}},\ }\bibfield  {title} {\enquote {\bibinfo {title} {{$\frac{1}{N}$
  expansion for the Kondo lattice}},}\ }\href
  {https://doi.org/10.1103/PhysRevB.28.5255} {\bibfield  {journal} {\bibinfo
  {journal} {Phys. Rev. B}\ }\textbf {\bibinfo {volume} {28}},\ \bibinfo
  {pages} {5255} (\bibinfo {year} {1983})},\ \bibinfo {note} {[Erratum:
  \href{https://doi.org/10.1103/PhysRevB.29.2829}{Phys. Rev. B {\bf 29}, 2829
  (1984)}]}\BibitemShut {NoStop}%
\bibitem [{\citenamefont {Nozi\`eres}\ and\ \citenamefont
  {Blandin}(1980)}]{Nozieres:1980}%
  \BibitemOpen
  \bibfield  {author} {\bibinfo {author} {\bibfnamefont {P.}~\bibnamefont
  {Nozi\`eres}}\ and\ \bibinfo {author} {\bibfnamefont {A.}~\bibnamefont
  {Blandin}},\ }\bibfield  {title} {\enquote {\bibinfo {title} {{Kondo effect
  in real metals}},}\ }\href {https://doi.org/10.1051/jphys:01980004103019300}
  {\bibfield  {journal} {\bibinfo  {journal} {J. Phys. France}\ }\textbf
  {\bibinfo {volume} {41}},\ \bibinfo {pages} {193} (\bibinfo {year}
  {1980})}\BibitemShut {NoStop}%
\bibitem [{\citenamefont {Eichten}(1988)}]{Eichten:1987xu}%
  \BibitemOpen
  \bibfield  {author} {\bibinfo {author} {\bibfnamefont {E.}~\bibnamefont
  {Eichten}},\ }\bibfield  {title} {\enquote {\bibinfo {title} {{Heavy quarks
  on the lattice}},}\ }\href {https://doi.org/10.1016/0920-5632(88)90097-7}
  {\bibfield  {journal} {\bibinfo  {journal} {Nucl. Phys. Proc. Suppl.}\
  }\textbf {\bibinfo {volume} {4}},\ \bibinfo {pages} {170} (\bibinfo {year}
  {1988})}\BibitemShut {NoStop}%
\bibitem [{\citenamefont {Paredes}\ \emph {et~al.}(2005)\citenamefont
  {Paredes}, \citenamefont {Tejedor},\ and\ \citenamefont
  {Cirac}}]{Paredes:2003}%
  \BibitemOpen
  \bibfield  {author} {\bibinfo {author} {\bibfnamefont {B.}~\bibnamefont
  {Paredes}}, \bibinfo {author} {\bibfnamefont {C.}~\bibnamefont {Tejedor}}, \
  and\ \bibinfo {author} {\bibfnamefont {J.~I.}\ \bibnamefont {Cirac}},\
  }\bibfield  {title} {\enquote {\bibinfo {title} {{Fermionic atoms in optical
  superlattices}},}\ }\href {https://doi.org/10.1103/PhysRevA.71.063608}
  {\bibfield  {journal} {\bibinfo  {journal} {Phys. Rev. A}\ }\textbf {\bibinfo
  {volume} {71}},\ \bibinfo {pages} {063608} (\bibinfo {year} {2005})},\
  \Eprint {http://arxiv.org/abs/cond-mat/0306497} {arXiv:cond-mat/0306497}
  \BibitemShut {NoStop}%
\bibitem [{\citenamefont {Duan}(2004)}]{Duan:2004}%
  \BibitemOpen
  \bibfield  {author} {\bibinfo {author} {\bibfnamefont {L.-M}\ \bibnamefont
  {Duan}},\ }\bibfield  {title} {\enquote {\bibinfo {title} {{Controlling
  ultracold atoms in multi-band optical lattices for simulation of Kondo
  physics}},}\ }\href {https://doi.org/10.1209/epl/i2004-10115-8} {\bibfield
  {journal} {\bibinfo  {journal} {Euro. Phys. Lett.}\ }\textbf {\bibinfo
  {volume} {67}},\ \bibinfo {pages} {721--727} (\bibinfo {year} {2004})},\
  \Eprint {http://arxiv.org/abs/cond-mat/0310394} {arXiv:cond-mat/0310394
  [cond-mat.soft]} \BibitemShut {NoStop}%
\bibitem [{\citenamefont {Gorshkov}\ \emph {et~al.}(2010)\citenamefont
  {Gorshkov}, \citenamefont {Hermele}, \citenamefont {Gurarie}, \citenamefont
  {Xu}, \citenamefont {Julienne}, \citenamefont {Ye}, \citenamefont {Zoller},
  \citenamefont {Demler}, \citenamefont {Lukin},\ and\ \citenamefont
  {Rey}}]{Gorshkov:2010}%
  \BibitemOpen
  \bibfield  {author} {\bibinfo {author} {\bibfnamefont {A.~V.}\ \bibnamefont
  {Gorshkov}}, \bibinfo {author} {\bibfnamefont {M.}~\bibnamefont {Hermele}},
  \bibinfo {author} {\bibfnamefont {V.}~\bibnamefont {Gurarie}}, \bibinfo
  {author} {\bibfnamefont {C.}~\bibnamefont {Xu}}, \bibinfo {author}
  {\bibfnamefont {P.~S.}\ \bibnamefont {Julienne}}, \bibinfo {author}
  {\bibfnamefont {J.}~\bibnamefont {Ye}}, \bibinfo {author} {\bibfnamefont
  {P.}~\bibnamefont {Zoller}}, \bibinfo {author} {\bibfnamefont
  {E.}~\bibnamefont {Demler}}, \bibinfo {author} {\bibfnamefont {M.~D.}\
  \bibnamefont {Lukin}}, \ and\ \bibinfo {author} {\bibfnamefont {A.~M.}\
  \bibnamefont {Rey}},\ }\bibfield  {title} {\enquote {\bibinfo {title}
  {{Two-orbital SU(N) magnetism with ultracold alkaline-earth atoms}},}\ }\href
  {https://doi.org/10.1038/nphys1535} {\bibfield  {journal} {\bibinfo
  {journal} {Nature Phys.}\ }\textbf {\bibinfo {volume} {6}},\ \bibinfo {pages}
  {289--295} (\bibinfo {year} {2010})},\ \Eprint
  {http://arxiv.org/abs/0905.2610} {arXiv:0905.2610 [cond-mat.quant-gas]}
  \BibitemShut {NoStop}%
\bibitem [{\citenamefont {Bauer}\ \emph {et~al.}(2013)\citenamefont {Bauer},
  \citenamefont {Salomon},\ and\ \citenamefont {Demler}}]{Bauer:2013}%
  \BibitemOpen
  \bibfield  {author} {\bibinfo {author} {\bibfnamefont {Johannes}\
  \bibnamefont {Bauer}}, \bibinfo {author} {\bibfnamefont {Christophe}\
  \bibnamefont {Salomon}}, \ and\ \bibinfo {author} {\bibfnamefont {Eugene}\
  \bibnamefont {Demler}},\ }\bibfield  {title} {\enquote {\bibinfo {title}
  {{Realizing a Kondo-Correlated State with Ultracold Atoms}},}\ }\href
  {https://doi.org/10.1103/PhysRevLett.111.215304} {\bibfield  {journal}
  {\bibinfo  {journal} {Phys. Rev. Lett.}\ }\textbf {\bibinfo {volume} {111}},\
  \bibinfo {pages} {215304} (\bibinfo {year} {2013})},\ \Eprint
  {http://arxiv.org/abs/1308.0603} {arXiv:1308.0603 [cond-mat.str-el]}
  \BibitemShut {NoStop}%
\bibitem [{\citenamefont {Nishida}(2013)}]{Nishida:2013}%
  \BibitemOpen
  \bibfield  {author} {\bibinfo {author} {\bibfnamefont {Yusuke}\ \bibnamefont
  {Nishida}},\ }\bibfield  {title} {\enquote {\bibinfo {title} {{SU(3) Orbital
  Kondo Effect with Ultracold Atoms}},}\ }\href
  {https://doi.org/10.1103/PhysRevLett.111.135301} {\bibfield  {journal}
  {\bibinfo  {journal} {Phys. Rev. Lett.}\ }\textbf {\bibinfo {volume} {111}},\
  \bibinfo {pages} {135301} (\bibinfo {year} {2013})},\ \Eprint
  {http://arxiv.org/abs/1308.3208} {arXiv:1308.3208 [cond-mat.quant-gas]}
  \BibitemShut {NoStop}%
\bibitem [{\citenamefont {Nakagawa}\ and\ \citenamefont
  {Kawakami}(2015)}]{Nakagawa:2016}%
  \BibitemOpen
  \bibfield  {author} {\bibinfo {author} {\bibfnamefont {Masaya}\ \bibnamefont
  {Nakagawa}}\ and\ \bibinfo {author} {\bibfnamefont {Norio}\ \bibnamefont
  {Kawakami}},\ }\bibfield  {title} {\enquote {\bibinfo {title} {{Laser-Induced
  Kondo Effect in Ultracold Alkaline-Earth Fermions}},}\ }\href
  {https://doi.org/10.1103/PhysRevLett.115.165303} {\bibfield  {journal}
  {\bibinfo  {journal} {Phys. Rev. Lett.}\ }\textbf {\bibinfo {volume} {115}},\
  \bibinfo {pages} {165303} (\bibinfo {year} {2015})},\ \Eprint
  {http://arxiv.org/abs/1506.02947} {arXiv:1506.02947 [cond-mat.quant-gas]}
  \BibitemShut {NoStop}%
\bibitem [{\citenamefont {Nakagawa}\ \emph {et~al.}(2018)\citenamefont
  {Nakagawa}, \citenamefont {Kawakami},\ and\ \citenamefont
  {Ueda}}]{Nakagawa:2018}%
  \BibitemOpen
  \bibfield  {author} {\bibinfo {author} {\bibfnamefont {Masaya}\ \bibnamefont
  {Nakagawa}}, \bibinfo {author} {\bibfnamefont {Norio}\ \bibnamefont
  {Kawakami}}, \ and\ \bibinfo {author} {\bibfnamefont {Masahito}\ \bibnamefont
  {Ueda}},\ }\bibfield  {title} {\enquote {\bibinfo {title} {{Non-Hermitian
  Kondo Effect in Ultracold Alkaline-Earth Atoms}},}\ }\href
  {https://doi.org/10.1103/PhysRevLett.121.203001} {\bibfield  {journal}
  {\bibinfo  {journal} {Phys. Rev. Lett.}\ }\textbf {\bibinfo {volume} {121}},\
  \bibinfo {pages} {203001} (\bibinfo {year} {2018})},\ \Eprint
  {http://arxiv.org/abs/1806.04039} {arXiv:1806.04039 [cond-mat.quant-gas]}
  \BibitemShut {NoStop}%
\bibitem [{\citenamefont {Toublan}\ and\ \citenamefont
  {Kogut}(2003)}]{Toublan:2003tt}%
  \BibitemOpen
  \bibfield  {author} {\bibinfo {author} {\bibfnamefont {D.}~\bibnamefont
  {Toublan}}\ and\ \bibinfo {author} {\bibfnamefont {J.B.}\ \bibnamefont
  {Kogut}},\ }\bibfield  {title} {\enquote {\bibinfo {title} {{Isospin chemical
  potential and the QCD phase diagram at nonzero temperature and baryon
  chemical potential}},}\ }\href
  {https://doi.org/10.1016/S0370-2693(03)00701-9} {\bibfield  {journal}
  {\bibinfo  {journal} {Phys. Lett. B}\ }\textbf {\bibinfo {volume} {564}},\
  \bibinfo {pages} {212--216} (\bibinfo {year} {2003})},\ \Eprint
  {http://arxiv.org/abs/hep-ph/0301183} {arXiv:hep-ph/0301183} \BibitemShut
  {NoStop}%
\bibitem [{\citenamefont {Son}\ and\ \citenamefont
  {Stephanov}(2001)}]{Son:2000xc}%
  \BibitemOpen
  \bibfield  {author} {\bibinfo {author} {\bibfnamefont {D.T.}\ \bibnamefont
  {Son}}\ and\ \bibinfo {author} {\bibfnamefont {Misha~A.}\ \bibnamefont
  {Stephanov}},\ }\bibfield  {title} {\enquote {\bibinfo {title} {{QCD at
  Finite Isospin Density}},}\ }\href
  {https://doi.org/10.1103/PhysRevLett.86.592} {\bibfield  {journal} {\bibinfo
  {journal} {Phys. Rev. Lett.}\ }\textbf {\bibinfo {volume} {86}},\ \bibinfo
  {pages} {592} (\bibinfo {year} {2001})},\ \Eprint
  {http://arxiv.org/abs/hep-ph/0005225} {arXiv:hep-ph/0005225} \BibitemShut
  {NoStop}%
\end{thebibliography}%
\end{document}